\documentclass{aa}
\usepackage{lineno}
\usepackage{graphicx}
\usepackage{soul}
\usepackage{dcolumn}
\usepackage{amsmath}
\usepackage{natbib}
\usepackage{slashed}
\usepackage{bm}
\usepackage{xcolor}
\usepackage[normalem]{ulem}

\bibpunct{(}{)}{;}{a}{}{,}
\usepackage{amsmath}
\usepackage{natbib}
\usepackage{slashed}
\bibpunct{(}{)}{;}{a}{}{,}

\title{Adiabatic evolution of hadronic neutron stars }

\author{
Domenico Logoteta\inst{1} 
\and Francesco Zeno Constanzo\inst{2} 
\and Ignazio Bombaci\inst{3,4} 
Sebastiano Bernuzzi\inst{5}
Albino Perego\inst{6,7}
David Radice\inst{8,9,10}
}
          
\institute{INFN, Sezione di Milano Bicocca, Piazza della Scienza, 3 20126 Milan, Italy 
\and Università di Ferrara, Via della Ghiara, 36, 44121 Ferrara, Italy
\and Università di Pisa, Largo Bruno Pontecorvo 3, 56127 Pisa, Italy  
\and INFN, Sezione di Pisa, Largo Bruno Pontecorvo 3, 56127 Pisa, Italy
\and Theoretisch-Physikalisches Institut, Friedrich-Schiller-Universit{\"a}t Jena, 07743, Jena, Germany
\and Dipartimento di Fisica, Università di Trento, Via Sommarive 14, I-38123 Trento, Italy
\and INFN-TIFPA, Trento Institute for Fundamental Physics and Applications, via Sommarive 14, I-38123 Trento, Italy
\and Institute for Gravitation and the Cosmos, The Pennsylvania State University, University Park, PA 16802, USA
\and Department of Physics, The Pennsylvania State University, University Park, PA 16802, USA
\and Department of AStronomy \& Astrophysics, The Pennsylvania State University, University Park, PA 16802, USA
}

\date{Received .........;  accepted ......}
\abstract 
{}
{
In this paper we compute several structural properties of static as well as rotating neutron stars considering both the case of rigid and differential rotation. 
To this aim we use a finite temperature $\beta$-stable equation of state (EOS) extending the zero-temperature hadronic EOS derived by Fortin et al. \citep{fortin17}. We take into account the possible formation of hyperons, as well as the presence of a neutrino trapped component. 
Our goal is to devise an approximate scheme to provide insights into the long-term evolution of a protoneutron star (PNS) formed after a core-collapse supernova (CCSNe) or a remnant possibly resulting from a binary neutron star merger (BNSM).  
} 
{The EOS is derived using a relativistic mean-field quantum many-body theory within the Hartree approximation. The coupling constants are determined using existing experimental data on hypernuclei and hypernuclear matter. This new set of couplings better accounts for uncertainties in the hyperon single-particle potentials in hadronic matter.  
Stationary General Relativity (GR) configurations are computed numerically for the novel EOS.}
{Our EOS models are able to reproduce the typical features of both cold neutron stars and PNSs, for both nucleonic and hyperonic matter compositions.  
In particular, our EOS models are consistent with present measured neutron star masses: $M = 2.01 \pm 0.04 \, M_{\odot}$ and $M = 2.14^{+0.20}_{-0.18} \, M_{\odot}$ corresponding to the neutron stars in PSR J0348+0432 and PSR J0740+6620, respectively. In addition, our EOSs predict values of the dimensionless tidal deformability parameter that are in agreement with the estimates provided by the analysis of the GW170817 event. Our study of thermal contributions in both static and rotating neutron stars shows that these effects act differently depending on the matter composition, as well as on the specific system considered, namely whether it is static or rotating.
Finally, our evolutionary scheme provides a simple way to interpret the long-term evolution of a remnant formed after a BNSM or of a PNS resulting from a CCSN. Our study represents a useful tool for complementing the information provided by dynamical numerical simulations, which are currently unable to cover the full timescales required for the analysis of these high-energy astrophysical phenomena. }  
{}


\titlerunning{Adiabatic evolution of hadronic neutron stars}

\authorrunning{Logoteta et al.}

\begin{document}

\maketitle
\nolinenumbers

\section{Introduction} 

The study of high-energy astrophysical systems, such as core-collapse supernovae (CCSNe) \citep[e.g.,][]{burrows_review_2013,janka_review_2016}, protoneutron stars (PNSs) \citep[e.g.,][]{prak97,roberts_PNS_2017,Janka2025,Boccioli2024,Bakir2026} 
, and binary compact star mergers involving at least one neutron star \citep[e.g.,][]{baiotti17,Shibata_review_2019,Radice_review_2020}, represents a formidable testbed for our understanding of gravity, nuclear physics, and particle physics. The extreme thermodynamical conditions realised in these astrophysical phenomena make their theoretical description highly challenging, and numerical relativity is currently the most appropriate approach to model the strong field dynamics that characterized these events.
The baryonic density ($n_B$) and temperature ($T$) values required by numerical simulations of CCSNe and compact binary mergers span at least $10$ orders of magnitude for $n_B$, extending up to several times nuclear saturation density, $n_0 = 0.16~\mathrm{fm^{-3}}$ 
(corresponding to $\sim 2.6 \times 10^{14}\mathrm{g~cm^{-3}}$), and $0\ {\rm MeV}< T \lesssim 100\ {\rm MeV}$. 
The scenario is further complicated by the fact that the actual matter composition, especially at very high densities and temperatures, is not precisely known; therefore, existing models rely on specific assumptions driven more by plausibility arguments and/or exploratory purposes than by a well-established theory.

In recent years, the detection of gravitational waves (GWs) from binary neutron star mergers (BNSMs) \citep{gw5, gw6} has provided a strong boost to research on dense matter physics. Gravitational-wave astronomy has indeed opened up the possibility of exploring matter under extreme conditions in regimes that are inaccessible to ground-based experiments.
Gravitational waves detected from the inspiral phase of BNSMs, especially when combined with robust lower limits on the maximum mass of neutron stars (NSs) \citep{Fonseca2021, Antoniadis2013, demo2010}, can set quite stringent constraints on the EOS of cold (i.e., $T = 0$) nuclear matter 
up to  $(2-3) \ n_0$ \citep{gw170817_eos} 


When combined with information extracted from the electromagnetic counterparts of GW events, inspiral signals can provide truly multimessenger constraints on the behavior of ultra-dense matter \citep[see e.g.,][]{Radice2018,Radice2018b,Breschi21, Breschi24, Pang23, Huth22}.  
Additionally, the post-merger and the post-bounce GW signals, generated by the remnant formed in a BNSM or in a CCSN event, and anticipated to be detected by third generation GW telescopes \citep[see e.g.,][]{Abac26}, have the potential of disclosing by themselves the properties of hot ($T \ge 10 \, \mathrm{MeV}$) and dense ($n \ge 2 n_0$) stellar matter.

In particular, post-merger GWs could carry information about the possible appearance of new particle species (e.g., hyperons) or new phases of matter (e.g., deconfined quark matter) \citep[see e.g.,][]{shibata2005,BJ2012,Hotokezaka2013,Bernuzzi2015,
RT2016,Bernuzzi2016,Bauswein2019,Most2020,Avi21}. Note in addition that the combined mass-radius measurements provided by the  Neutron Star Interior Composition Explorer (NICER) experiment \citep{nicer1, nicer2} set other independent constraints on the high-density EOS. 

The purpose of the present work is to provide a simple but robust scheme to describe the long-term evolution of a PNS or of a remnant resulting from a BNSM.
To this end, we describe the key stages of the evolution of the compact object as a sequence of snapshots, each corresponding to an equilibrium configuration computed at a given time. The dynamical and thermodynamic conditions that characterise each stage of the evolution are crucial for our analysis and will be discussed in detail in the following. A particularly important stage is the initial one, in which we assume that the compact object is differentially rotating. This hypothesis, now confirmed by accurate dynamical numerical simulations, has been widely investigated in the literature \citep{KEH89,cook92,cook94,sterg95,bau00,bon03,galeazzi12,ansorg1,ansorg2,ansorg3,rosinska17,espino19,Zhou19,Szk19,bozzola18,
Morrison04,Pili15,Bucciantini11,sold21,pass20,Camelio21}, and several codes have been developed to carry out this line of research. One of the main differences among these codes concerns the techniques adopted to solve the Einstein equations; finite-difference \citep[e.g.][]{sterg95} or spectral methods \citep[e.g.][]{ansorg1,ansorg3} are typically employed.
A comparison among the various existing codes for rapidly rotating stars, as well as their relative accuracy, can be found, for instance, in \citet{sterg03}. In our study, we employ the code presented in \citet{sterg95}, which, following \citet{KEH89}, rewrites the system of partial differential equations describing the Einstein equations, as a system of integral equations by means of the Green-function method. The resulting equations are then solved iteratively, starting from a suitable initial guess for the unknown functions, namely the metric potentials, the energy density, and the angular velocity.
The code was successfully used in several works to study remnant evolution \citep{kaplan14} and to search for quasi-universal relations among gravitational mass, rest mass, and angular momentum in differentially rotating neutron stars \citep{weih18,bozzola18}.
We also note that, in recent years, some codes have successfully included the effects of magnetic fields in calculations of rotating neutron star structure \citep{sold21,sold20}. A further recent development concerns a numerical code capable of studying equilibrium models with non-barotropic EOSs \citep{Camelio21}.

Concerning the composition of matter, in the present work we model the cold neutron star core in two different ways:
\begin{enumerate}
\item[i)] as a uniform charge-neutral fluid of neutrons, protons, electrons, and muons;  \\
\item[ii)] as a uniform charge-neutral fluid of neutrons, protons, all the hyperons of the baryonic octect ($\Lambda$, $\Sigma^{+,-,0}$ $\Xi^{0,-}$), electrons and muons.
\end{enumerate}
In order to describe the evolution of the early stages of a remnant formed in CCSNe and BNSM, cold EOSs are additionally extended at finite temperature including photons, and antiparticles.
Each EOS is calculated in $\beta$-equilibrium, a condition rapidly achieved by the compact object formed in CCSNe and BNSM.
In addition when describing specific scenarios in which thermal effect are expected to be relevant, we consistently include the effect of neutrino trapping in our matter description. In the early stages ($t<10 {\rm s}$) of the evolution of our astropysical systems, densities and temperatures are such that the neutrino mean free path is indeed much smaller than the typical radius of the star \citep{vidana03,vidana22}. 
These conditions are expected to hold up to $\sim 10 {\rm s}$ after the formation of a PNS or of a BNSM remnant, at least until the gravitational collapse to a black hole can occur.   

The paper is organised as follows. In Section \ref{rmf}, we describe the derivation of the EOS used in the present work. In Section \ref{results_EOS}, we present the results concerning the EOS for stellar matter, showing its behavior and the matter composition under several different physical scenarios.
Section \ref{results_evolution} is devoted to astrophysical applications of the EOS, namely the calculation of PNS structures using our EOSs in various cases: static configurations, rigid rotation, and differential rotation. We then apply these results to describe the evolution of a remnant produced by a BNSM or a PNS resulting from a CCSN.
In the final section, we summarise our results and outline the main conclusions of the present work.

\section{Relativistic mean field many-body theory}
\label{rmf}


In this work, we employ a finite-temperature relativistic mean-field (RMF) model to derive the EOS of hadronic matter. The nucleonic part of this model (hereafter DDME2) was developed in \citet{lagaris} by adjusting the coupling constants of the RMF Lagrangian of the earlier DD2 EOS to improve the description of certain properties of both spherical and deformed nuclei. The DDME2 model provides a good description of nuclear matter properties around saturation density, as can be seen in Table \ref{tab1_N}, where we report various nuclear matter parameters calculated at saturation density. 
Specifically besides $n_0$  we report the corresponding value of the energy per particle of symmetric nuclear matter, incompressibility, symmetry energy, and the slope of the symmetry energy. All these values are in good agreement with current experimental constraints for these quantities \citep{Lattimer14}. It should be noted that the DDME2 model, like DD2, has been constructed with density-dependent coupling constants. This density dependence was originally introduced in \citet{DD2} to develop a phenomenological RMF model capable of reproducing the behaviour observed in certain microscopic Dirac–Brueckner–Hartree–Fock calculations.
The DDME2 model was then extended in \citet{fortin17} by including the possibility of the formation of hyperons. The Lagrangian we employ in the present work reads: 

\begin{table}
\centering
\begin{tabular}{cccccc}
\hline 
$n_{0}$ [fm$^{-3}$] & $(E/A)$ [{\rm MeV}] & $K_{\infty}$ [{\rm MeV}] & $E_{sym}$ [{\rm MeV}] & $L$ [{\rm MeV}] \\ 
\hline 
0.152 & -16.1 & 250.9 & 32.3 & 51.2 \\ 
\hline 
\end{tabular}
\caption{Saturation density $n_0$ (first column) and corresponding energy per nucleon $E/A$ for symmetric nuclear matter (second column); incompressibility $K_\infty$ (third column), symmetry energy $E_{sym}$ (fourth column) and its slope parameter $L$ (fifth) calculated at saturation density. }
\label{tab1_N}
\end{table}

\begin{eqnarray}
\mathcal{L}&=&\sum_{B} \overline{\psi}_{B} (  \slashed{D}_{B} 
- m^*_{B}) \psi_{B} 
+ \sum_{l = e, \mu} \overline{\psi}_{l}  ( \partial \!\!\! / - m_{l}) \psi_{l} \cr
&+& \frac{1}{2}(\partial_{\mu} \sigma \partial^{\mu} \sigma - m^2_{\sigma} \sigma^2)  \cr
&-&\frac{1}{4} \Omega_{\mu \nu} \Omega^{\mu \nu} + \frac{1}{2} m_{\omega}^2 \omega_{\mu}\omega^{\mu} 
- \frac{1}{4} P_{\mu \nu} P^{\mu \nu} + \frac{1}{2} m_{\rho}^2 \rho_{\mu}\rho^{\mu}  \cr
&+& \frac{1}{2}(\partial_{\mu} \sigma^{*} \partial^{\mu} \sigma^{*} - m^2_{\sigma^{*}} \sigma^{*^2}) - \frac{1}{4} \Phi_{\mu \nu}  \Phi^{\mu \nu} + \frac{1}{2} m_{\phi}^2 \phi_{\mu}\phi^{\mu} \quad \;,
\label{lagr}
\end{eqnarray}
where we define:
\begin{eqnarray}
 \slashed{D}_{B} 
&=& \gamma_{\mu}(i \partial^{\mu} - g_{\omega B} \omega^{\mu} - g_{\phi B} \phi^{\mu} - g_{\rho B} \vec{\tau}_B \cdot \vec{\rho}\,^{\mu})  \;, \\
m^*_B &=& m_B - g_{\sigma B} \sigma - g_{\sigma^* B} \sigma^* \,, \\
 \Omega_{\mu \nu} &=& \partial_{\mu} \omega_{\nu} - \partial_{\nu} \omega_{\mu} \;, \\
 P_{\mu \nu} &=& \partial_{\mu} \rho_{\nu} - \partial_{\nu} \rho_{\mu} \;, \\
 \Phi_{\mu \nu} &=& \partial_{\mu} \phi_{\nu} - \partial_{\nu} \phi_{\mu} \;.
\end{eqnarray}
The masses of the various hyperons have been set as follows: $m_{\Lambda} = 1115.683\ \mathrm{MeV}$, $m_{\Sigma^+} = 1189.37\ \mathrm{MeV}$, $m_{\Sigma^0} = 1192.642\ \mathrm{MeV}$, $m_{\Sigma^-} = 1197.45\ \mathrm{MeV}$, $m_{\Xi^0} = 1314.86\ \mathrm{MeV}$, and $m_{\Xi^-} = 1321.71\ \mathrm{MeV}$. 
We note that, in recent years, the additional contribution of the $\Delta$ quartet was also investigated in several studies, showing that $\Delta$ formation can significantly modify the hadronic matter composition and, consequently, the global properties of the resulting stellar structure \citep{drago14,dell,Parmar14}. However, these calculations are strongly affected by the uncertainties in the nucleon-$\Delta$ couplings, and we leave this exploration to a dedicated future work.  

\subsection{Couplig constants}
In order to fix the coupling constants involved in the hyperonic sector, we have followed the strategy devised in \citet{fortin17}.
We have considered two possibilities: constant couplings and density-dependent (DD) coupling constants. 

We first describe the former.
It is well known that, while the nucleon–nucleon (NN) interaction is well constrained thanks to a large body of experimental data (for example, scattering data), the hyperon–nucleon (YN) interaction is subject to far fewer experimental constraints. If we additionally consider the hyperon–hyperon (YY) interaction, the situation becomes even more uncertain, as no scattering data exist. An alternative way to obtain information on the YN and YY interactions is through the study of hypernuclei. 
At present, several experimental measurements are available on the $\Lambda$-separation energies of the so called single $\Lambda$-hypernuclei. In \ \citet{fortin17} these information were used to fix the coupling between the $\Lambda$-hyperon and the 
$\sigma$-meson.
Concerning the other hyperons, namely $\Sigma$ and $\Xi$, we have fixed the corresponding couplings with the $\sigma$-meson using the expected value of their single particle energy $U_{H}$ (where $H=\Sigma,\ \Xi $) in symmetric nuclear matter at saturation density:
\begin{equation}
g_{\sigma H} = \frac{-U_{H} + g_{\omega H}\, \omega(n_{\rm sat})}{\sigma(n_0)} \;.
\label{eqforcc}
\end{equation}
We have used $U_\Sigma = 0$ and $U_\Xi = -14\ \mathrm{MeV}$ for the $\Sigma$ and $\Xi$ hyperons, respectively. It should be noted that $U_\Lambda$ can also be obtained by extrapolating the $\Lambda$ separation energy in single-$\Lambda$ hypernuclei to the limit $A \to \infty$. On the other hand, the choice of $U_\Sigma = 0\ \mathrm{MeV}$ is somewhat arbitrary; in the literature, $U_\Sigma$ is typically assumed to lie in the range $-5\ \mathrm{MeV} \leq U_\Sigma \leq +30\ \mathrm{MeV}$. Indeed, it is still unclear whether $\Sigma$-hypernuclei are actually bound.
Concerning the couplings of the vector mesons, we adopt the {\rm SU(6)} symmetry. In addition, due to the limited available experimental data, following \citet{fortin17} we have considered only the coupling with $\Lambda$-hyperon for the hidden strange $\phi$ and $\sigma^*$ mesons. Therefore
we have: 
\begin{align} 
& g_{\omega N} = \frac{3}{2} g_{\omega \Sigma} = \frac{3}{2} g_{\omega \Lambda} = 3 g_{\omega \Xi} \;, \\
& g_{\rho N}   = 2 g_{\rho \Sigma} = g_{\rho \Xi} \;, \\
& - \frac{\sqrt{2}}{3} g_{\omega N} = g_{\phi \Lambda} \;.
\end{align} 
For the $\sigma^*$ meson, again using {\rm SU(6)} symmetry relations, we finally get: $g_{\sigma \Lambda} = 0.538 \, g_{\sigma N}$.

We now discuss the setup of the hyperonic coupling constants in the case of DD couplings. When considering the density-dependent parametrisation of the coupling constants, following \citet{fortin17}, we multiply the values of the constant couplings obtained at saturation density by the same density-dependent functions employed in the purely nucleonic sector \citep{lagaris,DD2}.
For the $\sigma^*$ and $\phi$ mesons, we use the density-dependent functions corresponding, respectively, to the $\sigma$ and $\omega$ mesons. This approach is intended to roughly mimic the effects of microscopic many-body forces between hyperons and nucleons \citep{logoteta_nny}. For extensive reviews on hyperons, see for instance 
\citet{logoteta_rev,hyp3}. 

\subsection{Derivation of the equations of state}
From the Lagrangian Eq.~\eqref{lagr} we have calculated the equation of motion of the various meson fields in Hartree approximation.
In order to close the system of equations, we must include the relations that take into account the conservation of the baryonic number and electric charge:  
\begin{align} 
&n_B = n_n + n_p + n_{\Lambda} + n_{\Sigma^+} + n_{\Sigma^0} + n_{\Sigma^-} + n_{\Xi^0} + n_{\Xi^-} \;, \\
&n_p + n_{\Sigma^+} = n_e + n_{\mu} + n_{\Sigma^-} + n_{\Xi^-} \;. 
\end{align}
Finally the chemical potential of the various species are related each other by the weak equilibrium conditions: 
\begin{align} 
&\mu_n = \mu_p + \mu_e \;,\\
&\mu_e = \mu_{\mu} \quad,\\
&\mu_n = \mu_{\Lambda} = \mu_{\Sigma^0} = \mu_{\Xi^0}\;, \\
&\mu_p = \mu_{\Sigma^+} \;,\\
&\mu_{\Xi^-} = \mu_{\Sigma^-} = \mu_n + \mu_e \;.
\end{align}
It should be noted that the latter equations must be modified in the presence of neutrino trapping (see Eq. \ref{chem-eq} below).
Once the above system of equations is solved, together with the equations of motion for the various fields, the calculation of the corresponding thermodynamical quantities is straightforward.

The extension of the formalism to finite temperature is also quite direct. We have followed the procedure outlined in \citet{DD2} for the DD2 model. We note that, since we aim to study the early stages of the evolution of a PNS or a remnant formed after a BNSM, we also consider the physical scenario in which neutrinos are trapped in the stellar matter, forming a Fermi sea. In neutrino-trapped matter, we impose the following additional conditions that fix the lepton fraction for each lepton family \citep{BL86,Bomb95}: 
\begin{align} 
\label{neu_033}
& Y_{L_e}=\frac{n_{\nu_e} + n_e }{n_B}= 0.33 \;, \\
& Y_{L_\mu}=\frac{n_{\nu_\mu} + n_\mu}{n_B} = 0   \;.
\end{align}
where $n_x$ represent the net (i.e. particles minus antiparticles) density abundance of the lepton species $x$. Everywhere needed, leptons are treated as an ideal Fermi gas in thermal and reaction equilibrium with the hadronic component. 
These relations were derived from the output of hydrodynamical simulations of CCSNe and have been widely used in the literature. However, recent numerical simulations of BNSM with improved neutrino transport schemes suggest smaller values of $Y_{L_e} \sim 0.05$–$0.1$ \citep{perego19,zappa23,foucart16,Radice23}. In the following calculations, we nevertheless adopt $Y_{L_e} = 0.33$ in order to emphasise the effect of neutrino trapping, and we comment in the final section on how our results would change if a smaller value of $Y_{L_e}$ were employed.
We also note that no net tauonic lepton number can appear due to the very high mass of the $\tau$ lepton, which effectively prevents their production. We further assume that the contribution of $\tau$ neutrinos could be neglected, since they will never represent the dominant species in terms of abundance and pressure contribution. In the following section, we present results for the particle composition at zero and finite temperature. In the latter case, we consider the physical scenario of stellar matter at constant entropy per baryon ($S/A = \mathrm{const}$) rather than the isothermal case. According to numerical simulations, this scenario provides a more realistic description of the thermal behaviour of matter inside a PNS or a remnant formed in a BNSM \citep{perego19}. 

\section{Results for EOS and matter composition}         
\label{results_EOS}

\begin{figure}
\centering
\includegraphics[scale=0.3]{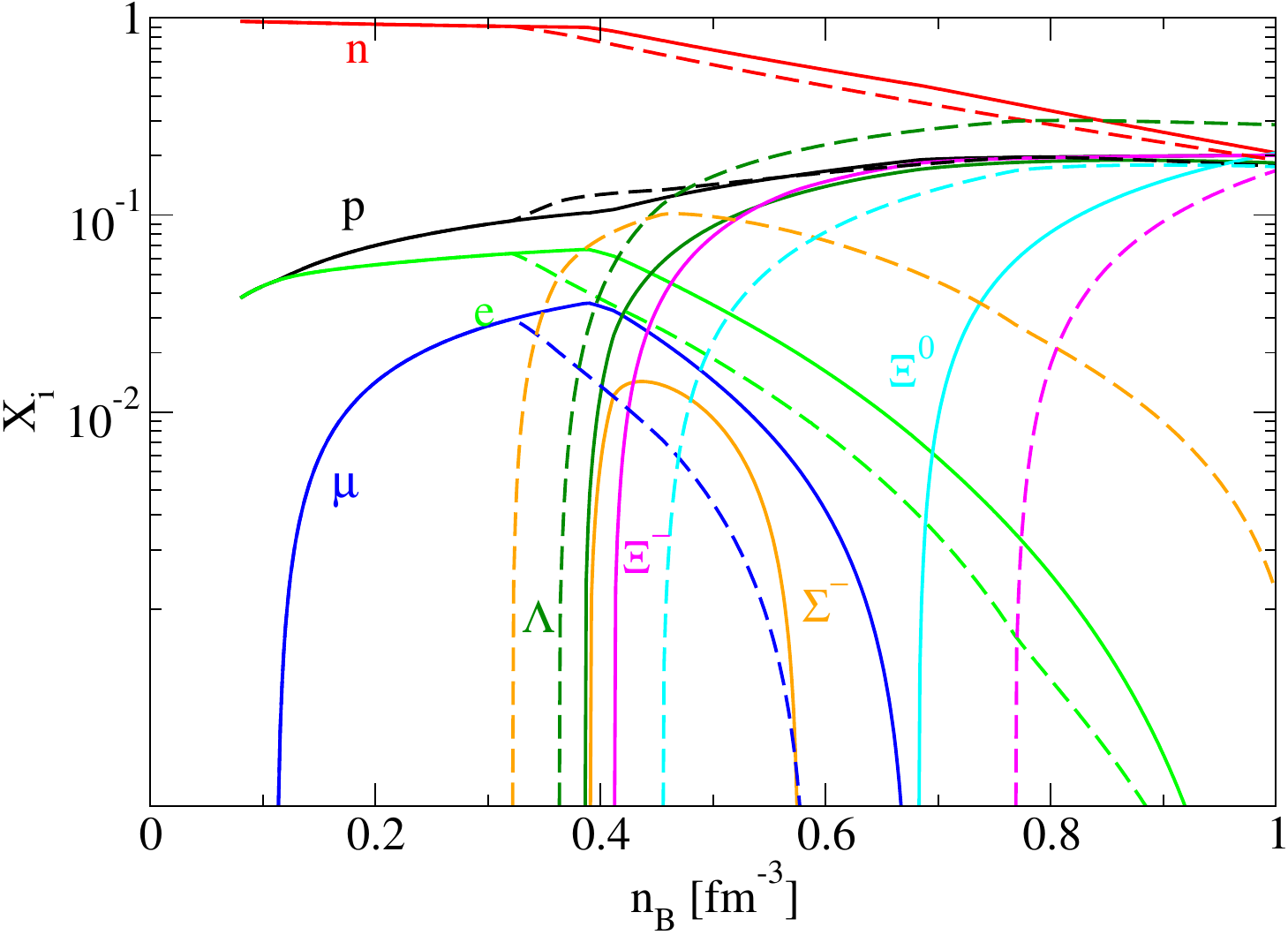}
\caption{Particles abundances ($X_i$) as function of the total baryonic numerical density for hyperonic matter. We show results for cold ($T = 0$) and neutrino-free matter, considering the case of density dependent (DD) coupling constants (dashed lines) and constant couplings (solid lines). }
\label{fig1}
\end{figure}

In this section, we present the main feature of the EOS obtained by considering the various possibilities concerning matter composition and thermodynamical conditions discussed previously. We do not report the zero-temperature composition for pure nucleonic matter, as it exhibits the typical features of $\beta$-stable nuclear matter, which are common to other RMF models as well as to microscopic calculations \citep{DD2,SFHO,BLh,BL}.

We begin by showing in Fig. \ref{fig1} the composition of cold matter in the case of hyperonic matter, considering density-dependent (DD) coupling constants (dashed lines) or constant couplings (solid lines).
A rather rich composition is obtained, and several hyperons of the baryonic octet appear. In the case of DD couplings, the first hyperon to appear is the $\Sigma^-$, with a threshold density of around $0.35~{\rm fm^{-3}}$. The other hyperons formed in $\beta$-stable matter according to the present model are $\Lambda$, $\Xi^-$, and $\Xi^0$, appearing in the listed order as the baryonic density increases. The largest hyperon fraction at the maximum density considered in our calculations is provided by the $\Lambda$ hyperon, while the $\Sigma^-$ fraction tends to vanish at high baryonic density. 
The decrease of the the $\Sigma^-$ population is initially due to the appearance of the charge neutral hyperons, $\Lambda$ and $\Xi^0$, and at larger density to the formation of the $\Xi^-$-hyperon. Being the latter charged, the system readjusts to keep charge neutrality and $\beta-$equilibrium conditions. Concerning the $\Lambda$ and the $\Xi^0$ hyperons, being electrically neutral particles, they do not enter in charge neutrality relation but only in the equation fixing of total baryon number density; their growth is therefore less constrained.

In the case of constant couplings, the first hyperon that appears is the $\Lambda$-one and its onset is slightly shifted to higher  baryonic densities compared with the DD case. The threshold densities for the appearance of the $\Lambda$ and $\Sigma^-$ are nearly the same ($\sim 0.38~{\rm fm^{-3}}$) and the $\Xi^-$-hyperon is formed just after. The threshold for the $\Xi^0$-hyperon is instead  shifted to higher densities around $\sim 0.78~{\rm fm^{-3}}$. Note that, for constant couplings, the $\Sigma^-$ population is substantially reduced compared with the DD coupling case.    
This is understood by noticing that in the case of constant couplings the onset density of the $\Lambda$, $\Sigma^-$, and $\Xi^-$ hyperons is very close; in addition charge neutrality and $\beta-$equilibrium favor the growth of the $\Xi^-$ population over the $\Sigma^-$ one.

\begin{figure}
\centering
\includegraphics[scale=0.3]{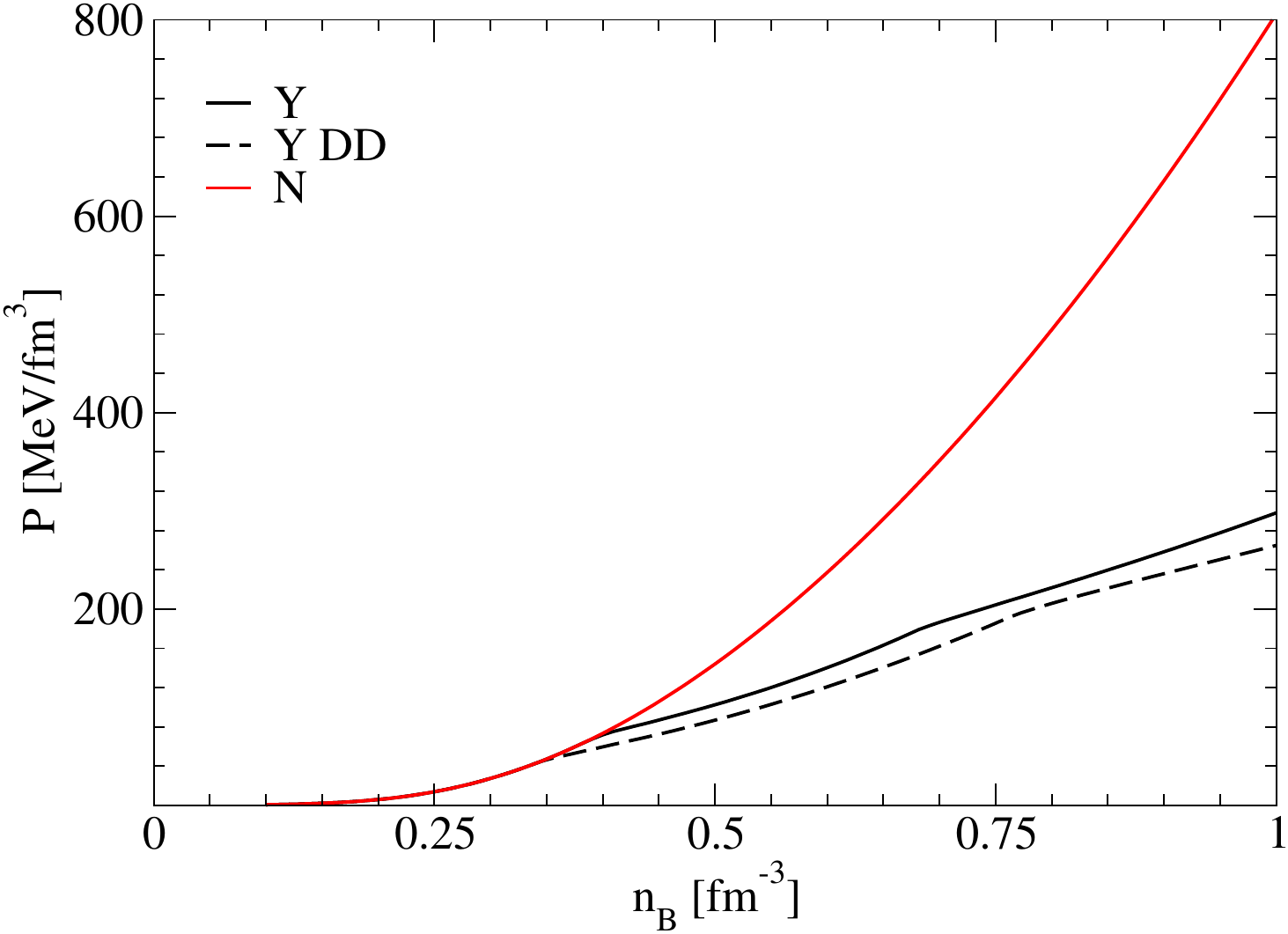}
\caption{EOS for nucleonic matter (red line), hyperonic matter with DD coupling constants (black dashed line) and with constant  couplings (black solid line). }
\label{fig2}
\end{figure}

To illustrate the overall effect of these particle abundances on the $\beta$-stable EOS, we show in Fig.\ \ref{fig2} the relation between pressure ($P$) and baryonic density for pure nucleonic matter (red line), hyperonic matter with density-dependent couplings (black dashed line), and hyperonic matter with constant couplings (black solid line).
Besides the well-known softening of the nuclear EOS due to the inclusion of hyperons, we note that the EOS with DD couplings is slightly softer than the EOS with constant couplings. Note also that the kinks appearing in the black lines, signalling the  different hyperon species.

We now turn to the case of abundances obtained at constant values of the entropy per baryon ($S/A$); this scenario is particularly important for describing the evolution of PNSs and remnants from BNSMs.

\begin{figure}
\centering
\includegraphics[scale=0.3]{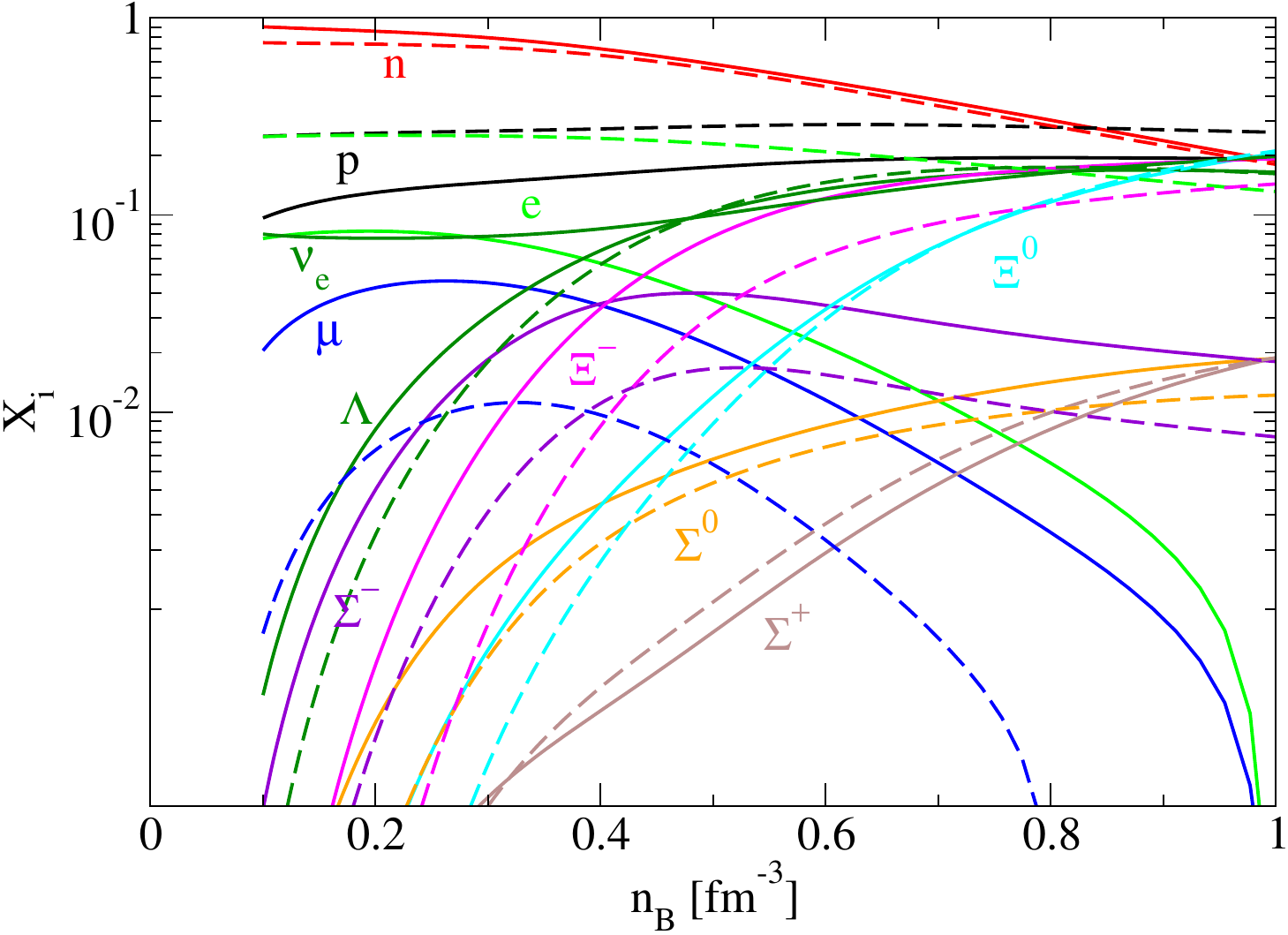}
\caption{Matter composition ($X_i$) as function of the baryonic density ($n_B$) in the case of a constant value of the entropy per baryon $S/A=2$ and considering neutrino trapped matter (dashed lines) and neutrino free matter (solid lines). We have adopted the case of constant couplings to set the hyperonic coupling constants.}
\label{fig3}
\end{figure}

\begin{figure}
\centering
\includegraphics[scale=0.3]{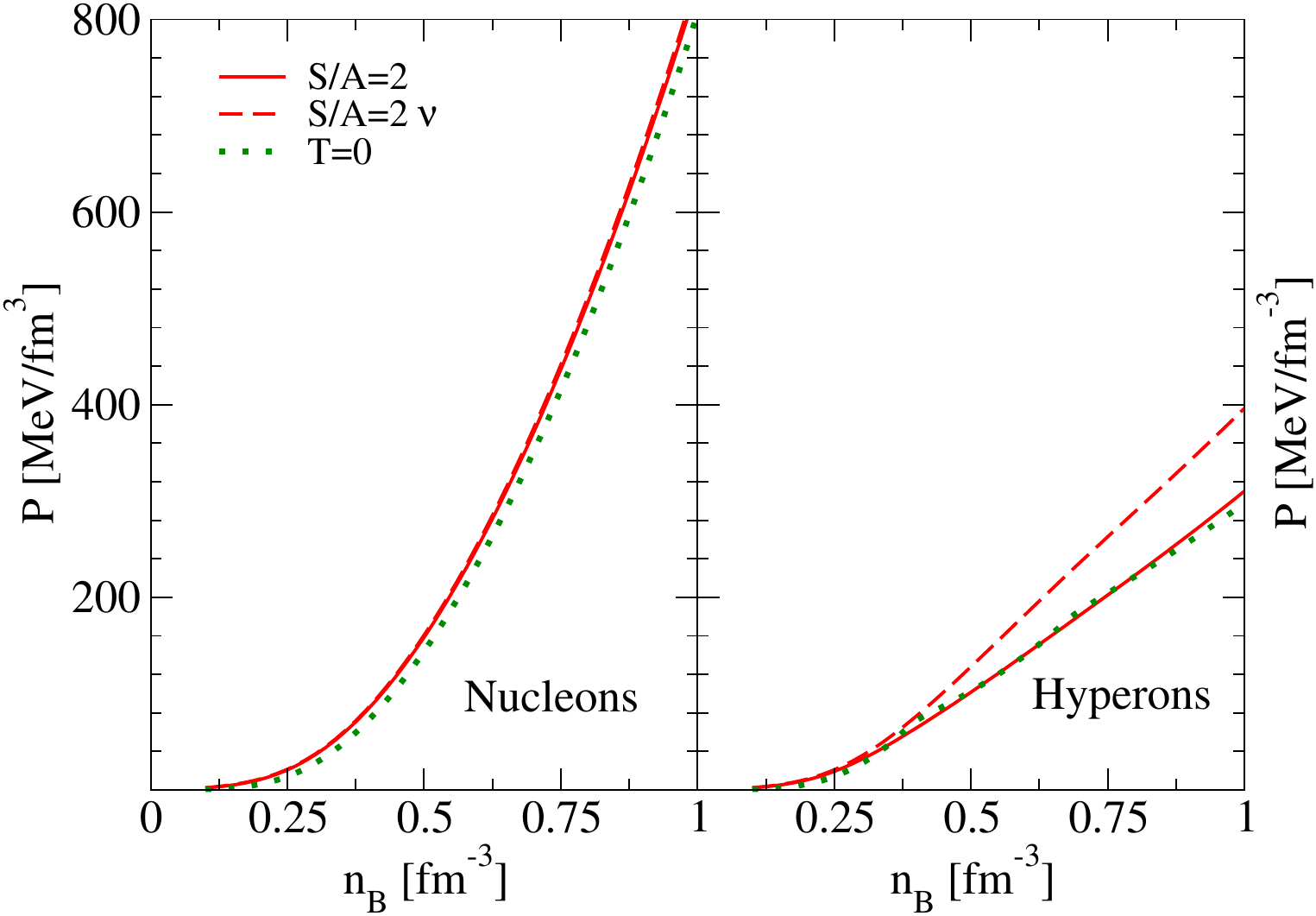}
\caption{Isoentropic EOSs for nucleonic matter (left panel), hyperonic matter (right panel) for $S/A=2$. The continuous line refer to $\nu$-free matter, the dashed one to $\nu$-trapped matter while the green dotted line refers to the $T=0$ $\nu$-free case. }
\label{fig4}
\end{figure}

In Fig.\ \ref{fig3}, we show the particle composition for $S/A = 2$ (in units of the Boltzmann constant $k_B$), including (solid  lines) or neglecting (dashed lines) the effect of neutrino trapping. In these calculations, we have considered the case of constant couplings to fix the hyperonic coupling constants.
The main aspect to highlight is that neutrino trapping reduces the strangeness content at a given baryonic density. This is confirmed by inspecting the corresponding EOSs, as shown in Fig. \ref{fig4}, where we report the nucleonic case with and without neutrino trapping (left panel) and the hyperonic matter case, again including or neglecting the effect of neutrino trapping.
It is interesting to note that, in nucleonic matter with trapped neutrinos, the electron chemical potential $\mu_e$ is increased by the electron neutrino chemical potential $\mu_{\nu_e}$. In this way, a larger fraction of electrons and consequently of protons, due to charge neutrality, is present at a given nuclear density. As a result, the EOS becomes softer, being more proton-rich. In hyperonic matter, the situation is somewhat more complex due to the presence of hyperons, which clearly affect the $\beta$-equilibrium reactions.
In general the chemical potential $\mu_i$ of a baryon species  $i$ in $\beta$-stable matter with trapped neutrinos can be written as:    
\begin{equation}
\mu_i = b_i \mu_n - q_i (\mu_e - \mu_{\nu_e})\; , 
\label{chem-eq} 
\end{equation} 
 where $b_i$ and $q_i$ are the baryon number and the electric charge (in units of the elementary electric charge) of the baryon species $i$, $\mu_n$ is the neutron chemical potential, and $\mu_e$ and $\mu_{\nu_e}$ are the chemical potentials of electrons and electron neutrinos respectively. 
 In the case of hyperonic matter using Eq.~\ref{chem-eq}, the chemical potential for 
the $\Sigma^-$ hyperon in neutrino trapped matter can be written as: 
\begin{equation}
         \mu_{\Sigma^-} =  \mu_n  + (\mu_e - \mu_{\nu_e})\; .
\label{chem-eq-sigma}
\end{equation} 
Since $(\mu_e - \mu_{\nu_e})$ in neutrino trapped matter is smaller than $\mu_e$ in neutrino free matter and since $\mu_n$ decreases too, as matter becomes more proton rich (see Fig.\ \ref{fig3}), 
the right hand side of Eq.(\ref{chem-eq-sigma}) decreases (at a given baryon density) when neutrinos are trapped.   
Thus the threshold density for the $\Sigma^-$ hyperons is shifted to a larger density and their abundance decreases with respect to the neutrino free case (Fig.\ \ref{fig3}). 

In the case of the $\Lambda$ hyperon (and similarly for other charge-neutral hyperons, $\Sigma^0$ and $\Xi^0$), the chemical equilibrium condition, Eq.~\ref{chem-eq}, gives $\mu_\Lambda = \mu_n$ both in neutrino-free and neutrino-trapped matter. At first glance, it might look that neutrino trapping has no effect on the chemical equilibrium condition and, therefore, on the threshold density for $\Lambda$ hyperons. 
However, this is not the case, because the chemical potentials of neutrons and $\Lambda$ hyperons do change when considering neutrino-trapped versus neutrino-less or neutrino-free matter. In fact, as can be seen in Fig.\ \ref{fig3}, neutrino trapping increases the proton fraction $X_p$ in $\beta$-stable nuclear matter, thereby reducing the neutron fraction $X_n$ and the neutron chemical potential $\mu_n$. Consequently, neutrino trapping shifts the threshold density for $\Lambda$ hyperons to higher values, in agreement with the results shown in the left panel of Fig.\ \ref{fig3}. 

As a result (Fig.\ \ref{fig4}, right panel), neutrino trapping produces a stiffening of the EOS of hyperonic matter.
These considerations are consistent with the findings of Refs. \citep{prak97,logoteta22} where the same qualitative results have been obtained employing different hadronic models. 

\section{Neutron star structure  }
\label{results_evolution}

In this section, we employ the EOSs discussed previously to calculate the structure of non-rotating and rotating neutron stars. This step is essential in order to describe the long-term evolution of PNSs and remnants from BNSMs, as we shall do in the final section of this work. 
We have used the publicly available code RNS to determine the structure of all equilibrium configurations. 
Since we do not perform any dynamical simulations in this paper, our goal is to identify and describe specific instants during the time evolution of the system that are relevant for characterising the final configuration reached after such long-term evolution. The latter can be either a stable neutron star or a black hole. Our scheme is as follows: we first describe the structure of the compact object formed immediately after a BNSM or a CCSNe. During this stage, the compact object is expected to rotate differentially, and the matter is hot and characterized by the presence of trapped neutrinos; we use an isoentropic EOS with $S/A = 2$ to mimic this stage.

Subsequently, differential rotation is lost due to several physical processes that are not yet completely understood \citep{Radice23}, and a state of rigid rotation may be approached.  
During this evolution, the matter remains hot. A possible complication arises because the timescale over which differential rotation is lost could be comparable to that of neutrino escape and the consequent deleptonisation of matter. This scenario may be particularly relevant for a remnant formed after a BNSM.

Given these uncertainties, we analyse both scenarios: one in which the loss of differential rotation occurs before neutrinos escape from the star, and one in which neutrinos and differential rotation are lost on the same timescale. The final configuration, if stable, is represented by a non rotating, cold ($T=0$), deleptonised neutron star, described by a solution of the TOV equations \citep{Tol34,OV39}; in all other cases, a black hole is formed.

\begin{figure}
\centering
\includegraphics[scale=0.3]{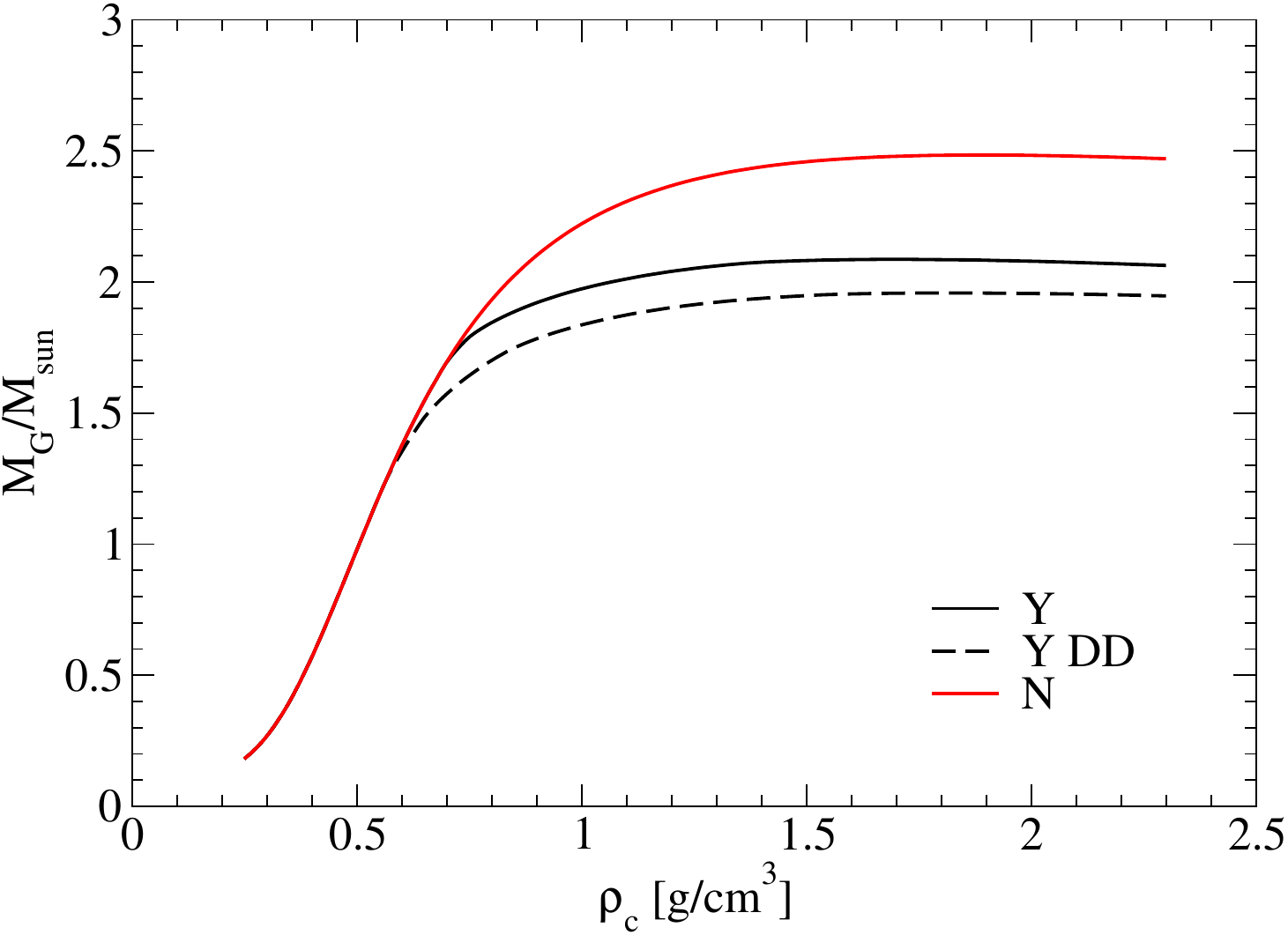}
\caption{Mass radius relation for nucleonic matter (red line), hyperonic matter with constant couplings (black line) and DD couplings (black dashed line). Energy density on the $x$-axis is reported in units of $10^{15}$ {\rm g/cm}$^3$.}
\label{fig5}
\end{figure}

Starting from the latter scenario, we show in Fig.\ \ref{fig5} the cold ($T=0$) TOV sequences obtained for nucleonic matter (red line) and hyperonic matter with an EOS described by constant (black continuous line) and density-dependent (black dashed line) coupling constants, respectively.

Note that, while the nucleonic EOS and the hyperonic EOS with constant couplings provide maximum gravitational masses above the $2 \ M_\odot$ limit \citep{demo2010,Antoniadis2013}, the EOS with DD couplings fails to satisfy this requirement. In the following, when discussing hyperonic matter, we will focus on the case of constant couplings. Accordingly, in the remainder of the text we drop the label "DD" when referring to a hyperonic EOS.

The various maximum-mass configurations (both baryonic and gravitational) for the three EOSs are reported in Tab. \ref{tab1}, along with the corresponding central energy density, $\rho_c$, and radius, $R$.
\begin{table}
\centering
\begin{tabular}{ccccc}
\hline 
 EOS    & $M_G$         & $M_B$         & $\rho_c $            & $R$ \\ 
     & [$M_{\odot}$] & [$M_{\odot}$] & $10^{15}$ [g/cm$^3$] & [km] \\ 
\hline 
N    & 2.484 & 2.996 & 1.895 &  11.997 \\ 

Y    & 2.086 & 2.417 & 1.708 &  12.421 \\

Y DD & 1.958 & 2.245 & 1.812 &  12.150 \\
\hline 
\end{tabular}
\caption{Maximum gravitational (first column) and baryonic mass (second column) configurations for the three cold EOSs discussed in the text. In the third and fourth columns we report the corresponding central energy density ($\rho_c$) in units of $10^{15}$ {\rm g/cm} $^3$ and radius ($R$). }
\label{tab1}
\end{table}
The mass-radius relations produced by our models are within the measure uncertainty of the recent constraints provided by the NICER  experiment. A measurement of the radius of PSR J0740+6620 was proposed in \ \cite{nicer1}, which set a limit of $R<13.7_{+2.6}^{-1.5}~{\rm km}$, and in \ \cite{nicer2}, where a limit of $R<12.39_{+1.3}^{-0.98}~{\rm km}$ was concluded. These results derive from two independent analyses of X-ray data obtained by NICER and the X-ray Multi-Mirror (XMM-Newton) observatory.
Specifically for a $1.4 \ M_\odot$ neutron star, our calculated radius is $13.13 {\rm km}$ for nucleonic matter and $13.12~{\rm km}$ for hyperonic matter. 
We have also calculated the tidal deformability parameter for a single star, which has been estimated from the GW event GW170817 \citep{gw5,gw170817_eos}. Our cold EOSs predict a tidal deformability, $\Lambda$, for a $1.4 \ M_\odot$ ($\Lambda_{1.4}$) neutron star of $653$ in the case of pure nucleonic matter and slightly smaller, $651$, in the case of hyperonic matter. These values are only slightly outside the $90$\% condifence level, $190^{+390}_{-120}$, reported in \citep{gw170817_eos}.

\subsection{No rotation}

Before discussing rotating neutron stars, which is one of the main objectives of this work, we examine how thermal effects modify the structure of static neutron stars in order to separate them from the rotational contributions, which we will introduce next.

\begin{figure}
\centering
\includegraphics[scale=0.3]{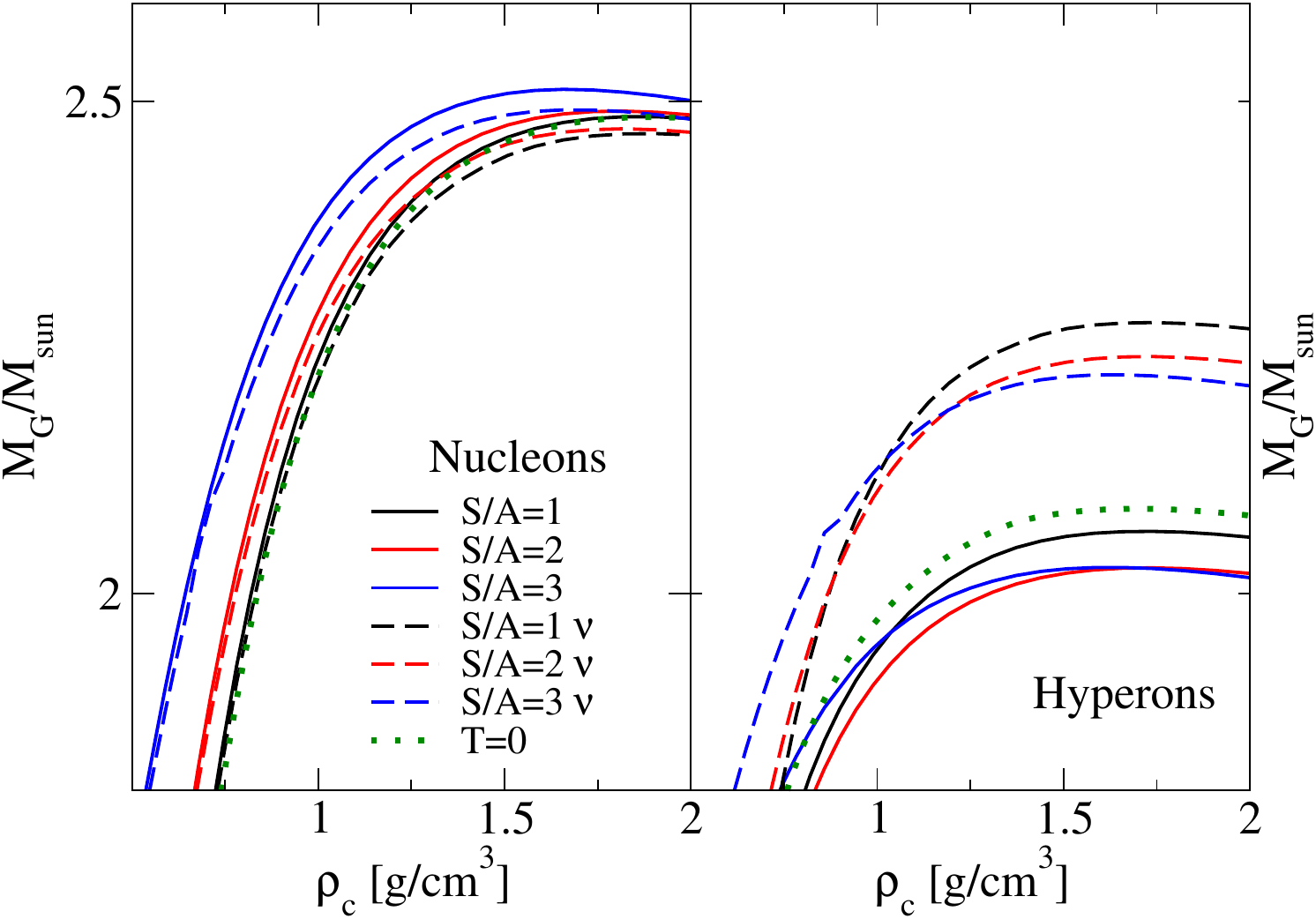}
\caption{Gravitational mass as function of the central energy density (in units of $10^{15}$ {\rm g/cm}$^3$) for nucleonic matter (left panel) and hyperonic matter (right panel) for several values of the entropy per baryon $S/A$. Dashed (solid) lines refer to neutrino trapped (free) matter. The green dotted line in both panels represents the zero temperature sequence.}
\label{fig5_new}
\end{figure}

In Fig.~\ref{fig5_new}, we show the gravitational mass ($M_G$, in units of solar mass) as a function of the central energy density, $\rho_c$, for several values of $S/A$, considering the effect of neutrino trapping (dashed lines) and neutrino-free matter (continuous lines). The green dotted line corresponds to the neutron star sequence at zero temperature. In the left panel, results are shown for pure nucleonic matter, while the right panel includes hyperons.

Focusing first on the nucleonic case, thermal effects lead to slightly larger gravitational neutron star masses than in the cold case. This is simply due to the thermal support, i.e. an incease in the pressure due to $T$. 
The presence of neutrino trapping reduces $M_G$ compared with the neutrino-free case. However, it is apparent that, in pure nucleonic matter, thermal effects, including neutrino trapping, produce only minor variations (of the order of $\sim 0.016 \ M_\odot$) relative to the cold case. Conversely, the maximum baryonic mass decreases as $S/A$ increases, indicating that neutron stars become less bound with increasing thermal effects. In \cite{kaplan14} similar calculations, with trapped neutrinos, were carried out employing different EOSs as well as making use of isothermal core's EOS supplemeted with an isothermal envelope at smaller temperature. The trend of our results agree with the ones reported in that work for the following EOSs 
: DD2, SFHo, SFHx. For these EOSs indeed, the thermal effects supplemented by neutrino trapping produces slightly smaller gravitational and baryonic masses compared to the cold case. In contrast, two other EOSs considered in \cite{kaplan14}, namely the LS and Shen EOSs, produce larger gravitational masses in the presence of neutrino trapping. Unfortunately we cannot compare results concerning the pure thermal effect without neutrino trapping because this case was not considered in \cite{kaplan14}.

Considering now the hyperonic case in neutrino-free matter, we observe a different trend: $M_G^{\rm max}$ indeed slightly decreases  with $S/A$ when hyperons are included. This is because thermal effects enhance the strangeness content at a given baryonic density, producing a small softening of the EOS and a consequent slight reduction in the maximum gravitational mass. Similarly, the baryonic mass decreases with increasing $S/A$. 

Another noteworthy aspect is that, while neutrino trapping in nucleonic matter reduces the maximum gravitational mass, in hyperonic matter it produces a stiffening of the EOS and consequently a larger $M_G^{\rm max}$ compared with the neutrino-free case. This effect produces also larger values of $M_B^{\rm max}$, and results from the previously discussed interplay of the chemical  potentials among the various species that induces a stiffening of the EOS.

\subsection{Rigid rotation}

We now present results for the case of rigid rotation, again employing our nucleonic and hyperonic EOSs. It is well known that uniform rotation can support a neutron star against gravitational collapse beyond the TOV limit. Those uniformly rotating neutron stars with a baryonic mass such that $M_B(\Omega) > M_B^{\rm max}(\Omega=0)$ are referred to as “supramassive,” where $M_B^{\rm max}(\Omega=0)$ is the maximum baryonic mass provided by the solution of the TOV equations.

For a fixed value of the central energy density, the gravitational mass that can be supported increases with rotational frequency until the system begins to lose mass from the equator along geodesics. This maximum frequency defines the so-called mass-shedding limit (also known as the Keplerian limit).  

\begin{figure}
\centering
\includegraphics[scale=0.3]{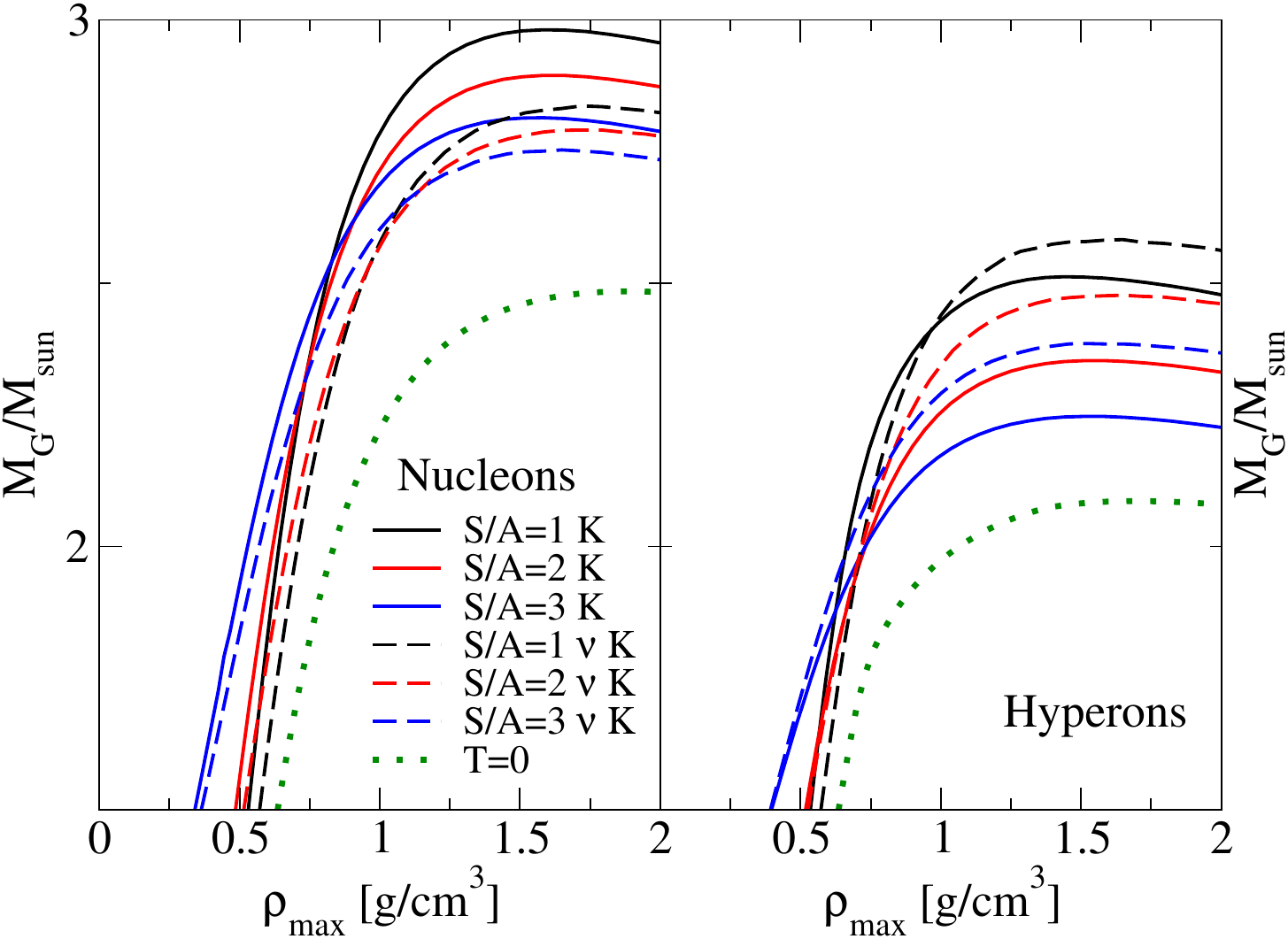}
\caption{In the left (right) panel we report the gravitational mass-central energy density relation at keplerian limit for nucleonic (hyperonic) matter for different values of $S/A$, considering neutrino free (solid lines) and neutrino trapped (dashed lines) matter. The green dotted line in both panels is the mass-central energy density relation for the cold nonrotating sequence. }
\label{fig6}
\end{figure}

In Fig. \ref{fig6}, we show the gravitational mass as a function of central energy density calculated at the Keplerian limit for nucleonic matter (left panel) and hyperonic matter (right panel) for three values of $S/A$. 
We also consider neutrino-free matter (solid lines) and neutrino-trapped matter (dashed lines). The green dotted line in both panels corresponds to the gravitational mass-central energy density relation for the cold, non-rotating sequence. 
Comparing the results of Fig.\ \ref{fig6} with the TOV equilibrium solutions, the gravitational masses are shifted to higher values due to the strong support provided by rapid rotation \citep{Datta98}. In general, we find that at the Keplerian limit, maximum masses are increased by approximately $20$\% compared with the static case. This occurs both at $T=0$ and at finite values of $S/A$. 

Several interesting aspects are worth noting. A key aspect is that, when assuming pure nucleonic matter independently on neutrino trapping, thermal effects act in the opposite way compared to the static case, lowering the maximum gravitational and baryonic masses while increasing $S/A$. This behavior was also found in some of the models reported in \citep{kaplan14}, in particular by those models constructed using cold envelopes in the EOS. 
The latter produce in general larger maximum gravitational masses than their finite-temperature counterparts. 
By contrast, we stress that in our EOS models the thermal contribution is entirely fixed by the value of $S/A$. 
Continuing with the comparison to the static case, we note that larger values of the entropy in rotating stars result in larger radii but also larger values of the central energy densities. Althought this may look not intuitive, it should be noticed that increasing the value of $S/A$ the keplerian limit is reached at smaller frequency. 
This in turn leads to larger values of the central energy density.

Table~\ref{tab2} summarizes the maximum-mass configurations, together with their corresponding central energy densities, radii, and Keplerian frequencies. For the isentropic models, we report only the representative $S/A=2$ case, since the results for the other entropy values do not provide any significant additional insight. 
We note that the maximum masses are systematically reached at smaller central energy densities compared to the static case, and the specific values decrease with increasing $S/A$. 

Moreover, these calculations confirm that, for nucleonic matter with neutrino trapping, the maximum masses supported are smaller than those for neutrino-free matter. In hyperonic matter, as in the static case, the opposite behaviour is observed. 

It is worth noting that, in the case of rigid rotation, similarly to the static case, the maximum energy density in a star coincides with the central energy density. However, this is not necessarily true for differentially rotating stars, as we discuss next. 
\begin{table}
\centering
\begin{tabular}{cccccc}
\hline 
 EOS & $M_G$         & $M_B$         & $\rho_{max} $        & $R_e$ & $\Omega_K$ \\ 
       & [$M_{\odot}$] & [$M_{\odot}$] & $10^{15}$ [g/cm$^3$] & [km] &  $10^3 \ {\rm rad} \ {\rm s}^{-1}$\\ 
\hline
N T=0          & 2.484 & 2.996 & 1.895 &  11.997 &  0\\
N T=0 K        & 2.964 & 3.602 & 1.594 &  15.966 & 10.557 \\
N S/A=2        & 2.486 & 2.895 & 1.820 &  12.828 & 0\\
N$\nu$ S/A=2   & 2.468 & 2.813 & 1.820 &  13.213 & 0\\
N S/A=2 K      & 2.872 & 3.333 & 1.657 &  16.879 & 9.136 \\
N$\nu$ S/A=2 K & 2.754 & 3.151 & 1.792 &  17.093 & 8.937 \\
Y T=0          & 2.062 & 2.416 & 1.707 &  12.421 &  0 \\
Y T=0 K        & 2.611 & 3.019 & 1.354 &  17.180 & 8.181 \\
Y S/A=2        & 2.020 & 2.264 & 1.737 &  13.262 &  0\\
Y$\nu$ S/A=2   & 2.234 & 2.496 & 1.737 &  13.808 &  0\\
Y S/A=2 K      & 2.331 & 2.627 & 1.581 &  18.034 & 7.532 \\
Y$\nu$ S/A=2 K & 2.437 & 2.745 & 1.649 &  15.517 & 7.461 \\
\hline 
\end{tabular}
\caption{
Maximum mass configuration for non-rotating cold ($T=0$) neutrino-free matter for different EOS models (first colum), 
maximum gravitational (second column) and baryonic mass (third column) configurations.  
We additional report for each configuration the corresponding central energy density (fourth column), the equatorial radius (fifth  column) and the Keplerian angular velocity (sixth column) in the case of rotating models. Rotating models are those with a label K in the first column.  }
\label{tab2}
\end{table}

\subsection{Differential rotation}

The final and particularly interesting scenario to discuss is that of stellar models supported by differential rotation. Differentially rotating equilibrium configurations have been shown to support masses even larger than the supramassive limit. Such configurations, with $M_B(\bar{A}) > M_B^{\rm max}(\Omega_K)$, are referred to in the literature as hypermassive neutron stars (HMNS), where $\bar{A}$ is a parameter defining the differential rotation profile, as discussed below.

The internal regions of the star can reach high rotation frequencies without the outer layers exceeding the Keplerian limit. The numerical calculations in this case are considerably more involved, and several remarks are in order. First, we note that the rotational profile implemented in the RNS code is not the most realistic one when compared with the outputs of dynamical simulations \citep{Kastaun2015,Han17,endrizzi18}.

We are aware that more sophisticated rotational profiles have been suggested \citep{galeazzi12}, such as those reported in  \citep{uryu17} and employed, for instance, in Refs. \cite{Fran22}; these profiles better reproduce the non-monotonic distribution of $\Omega$ observed in HMNS remnants formed in dynamical BNSM simulations. In our version of the RNS code, only the so-called $J$-constant law is currently implemented. Improved versions of RNS have been developed in Refs.\ \citep{bau17,pang21} to allow for more general rotational profiles.


The rotational profile we employ is parameterised as follows:
%
%
\begin{equation}
F(\Omega)=u^t u_\phi=A^2 (\Omega_c - \Omega) \;.
\end{equation}
Here, $F$ denotes the gravitationally redshifted angular momentum per unit rest mass and enthalpy, while $u^t$ and $u_\phi$ are components of the $4$-velocity. $\Omega$ is the angular velocity, and $A$ and $\Omega_c$ are parameters of the model. It is customary to introduce a rescaled parameter, $\hat{A} = A / r_e$, where $r_e$ is the equatorial radius.

The parameter $\hat{A}$ provides a measure of the degree of differential rotation and determines the length scale over which $\Omega$ varies within the star. In the limit $\hat{A} \to \infty$, uniform rotation is recovered, whereas in the limit $\hat{A} \to 0$, the specific angular momentum is constant throughout the star. In the following, we parametrise our equilibrium sequences in terms of the new parameter $\overline{A} = \hat{A}^{-1}$.

We also note that, as stated previously, in the cases of static and rigid rotation, the central energy density ($\rho_c$) coincides with the maximum energy density ($\rho_{\rm max}$) of the star. In the case of differential rotation, due to the non-uniform frequency profile, the maximum energy density may be displaced from the centre. When presenting results, we have chosen to plot various quantities against $\rho_{\rm max}$, which represents a physically meaningful quantity, rather than $\rho_c$, which is merely a parameter of the calculation.

Another important aspect concerns the definition of the mass-shedding limit for differentially rotating stars. In the case of rigid rotation, this limit is determined by the maximum (Keplerian) frequency beyond which the star begins to eject mass from the equator.  For differential rotation, this definition is not generally applicable, since different rotational profiles can be adopted, each with its own specific mass-shedding limit. It has been shown that, in the case of differential rotation, the Keplerian limit should be achieved when a certain quantity known as the rescaled shedding parameter vanishes \citep{ansorg1}:
%
\begin{equation}
\tilde{\beta}=\frac{\beta}{1+\beta} \;,
\end{equation}
where $\beta=-\frac{r_e^2}{r_p^2}\left(\frac{d z^2_b(\rho)}{d \rho^2}\right)_{r_e}$. 
In the above expressions, $r_p$ ($r_e$) denotes the polar (equatorial) radius, $\rho$ is the cylindrical radial coordinate, and $z_b$ is a function describing the shape of the stellar surface.

The quantity $\tilde{\beta}$ varies between $0$ and $1$, with several interesting limiting cases: $\tilde{\beta} = 0$ corresponds to the mass-shedding limit; when $\tilde{\beta} \to 1$, the star’s topology becomes toroidal; and for $\tilde{\beta} = 1/2$, the star is non-rotating and spherical. Unfortunately, it is not possible to reach the limit $\tilde{\beta} = 0$ using the RNS code.

To estimate the maximum mass supported by a given model, we have followed the strategy suggested in Refs.\ \citep{kaplan14,bau00}. We first fix a value of $\overline{A}$ and the central energy density, then perform several calculations for different values of the oblateness, defined as the ratio of polar to equatorial radius, $r_p / r_e$. The maximum baryonic mass as a function of oblateness provides our best estimate of the maximum mass supported for a given $\rho_c$. The corresponding gravitational mass is the value reported on the gravitational mass-maximum energy density plot.

We note, however, that the RNS code often fails to converge before mass detachment occurs, and therefore possibly before the true maximum mass of a given configuration is reached. This limitation implies that the maximum masses reported here should be interpreted as lower limits for the actual maximum masses of each model.


In the following, we consider two values of $\overline{A}$, namely $0.4$ and $1$. This choice is motivated by the findings of \ \citep{ansorg3}, which showed that for the adopted differential rotation profile (using a polytropic EOS with $\Gamma = 2$), there exists a critical value $\overline{A}_{\rm crit} \sim 0.7$: models with $\overline{A} > \overline{A}_{\rm crit}$ reach the mass-shedding limit via quasi-toroidal structures, whereas models with $\overline{A} < \overline{A}_{\rm crit}$ reach it via spheroidal structures. Our choice of $\overline{A} = 0.4$ and $1$ thus brackets this critical value.

Before presenting results for differentially rotating models, an important point to discuss concerns the stability of our equilibrium configurations. In the case of non-rotating, spherical stars, configurations are dynamically unstable against radial perturbations if $\frac{\partial M_G}{\partial \epsilon_c}<0$ which follows from the well known turning-point method \citep{sorkin82}. Although a general criterion for the dynamical stability of rotating stars does not exist, numerical simulations of uniformly rotating models suggest that the onset of dynamical instability occurs very close to the onset of secular instability \citep{shibata99}. In the following, we adopt the above criterion to select stable models. As an additional indicator of stability, we monitor the ratio $T/W$, where $T$ is the rotational kinetic energy and $W$ the gravitational binding energy. It has been shown \citep{Baiotti07,Manca07} that configurations with $T/W \gtrsim 0.25$ are dynamically unstable to non-axisymmetric perturbations, due to the onset of the so-called bar-mode instability \citep{offler15,Manca07}.

\begin{figure}
\centering
\includegraphics[scale=0.3]{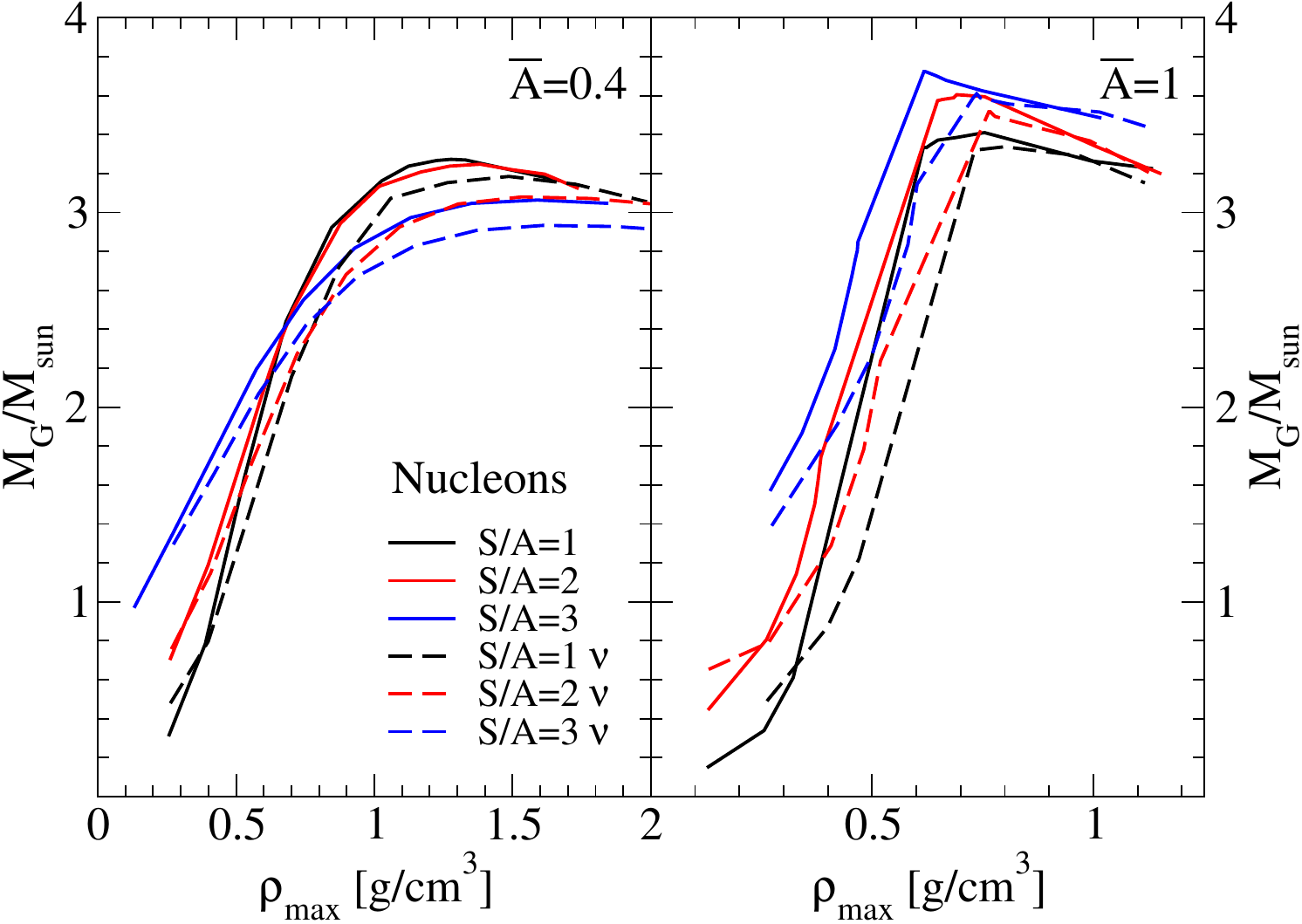}
\caption{Maximum gravitational mass as function of the maximum energy density for nucleonic matter considering two different  differential rotation rates $\bar{A}$: $0.4$ (left panel) and $1$ (right panel). Results are reported for various values of $S/A$ for neutrino free (solid lines) and neutrino trapped (dashed lines) matter.  }
\label{fig7}
\end{figure}

We begin by presenting in Fig.\ \ref{fig7} the sequences obtained for two different values of $\overline{A}$: $0.4$ (left panel) and $1$ (right panel), considering several values of $S/A$ and nucleonic matter both without (solid lines) and with (dashed lines) neutrino trapping. The qualitative behavior for $\overline{A}=0.4$ is quite similar to that observed in the case of rigid rotation (see Fig.\ \ref{fig6}). We emphasise that $\overline{A}=0.4$ corresponds to a low degree of differential rotation, which becomes zero in the limit $\overline{A} \to 0$. The maximum masses are systematically larger than in the rigid rotation case. As in the rigidly rotating models, maximum masses increase as $S/A$ decreases.

The equilibrium configurations obtained for $\overline{A}=0.4$ are all spheroidal, with the maximum energy density coinciding with  the centre of the star. The oblateness in these models never exceeds approximately $0.45$.

By contrast, the scenario for $\overline{A}=1$ exhibits significant differences. Irrespective of the presence of neutrinos, the maximum masses increase with increasing $S/A$. Several factors contribute to this behaviour. While for $\overline{A}=0.4$ we have $\rho_c \sim \rho_{\rm max}$, this is no longer generally true for $\overline{A}=1$. In this case, beyond a critical value of $\rho_{\rm max}$ along the equilibrium sequence, the location of $\rho_{\rm max}$ shifts from the centre to a circular ring at a finite distance from the star’s symmetry axis. In other words, for energy densities above a threshold that depends on $S/A$ and the matter composition, the configurations with $\overline{A}=1$ adopt a quasi-toroidal shape. For example, in the case of nucleonic, neutrino-free matter with $S/A=3$, this transition occurs at $\rho_{\rm max}=0.47 \times 10^{15}\ {\rm g/cm^3}$, while in neutrino-trapped matter it occurs at $\rho_{\rm max}=0.61 \times 10^{15}\ {\rm g/cm^3}$.

For $\overline{A}=1$, the oblateness near the maximum supported mass is generally smaller than for $\overline{A}=0.4$. However, the RNS code often fails to converge for oblateness values below $\sim 0.2$. Moreover, in the vicinity of the maximum mass, the ratio $T/W$ approaches the critical value of $0.25$, indicating proximity to the onset of bar-mode instability.

Another important point is that models with a higher degree of differential rotation exhibit significantly lower values of $\rho_{\rm max}$ compared with those with a low degree of differential rotation. The central angular velocities $\Omega_c$ for $\overline{A}=1$ are also substantially higher than those for $\overline{A}=0.4$, as shown in Tables \ref{tab3} and \ref{tab4}. The larger rotation rates provide stronger centrifugal support, which favours configurations with higher gravitational masses.

\begin{figure}
\centering
\includegraphics[scale=0.3]{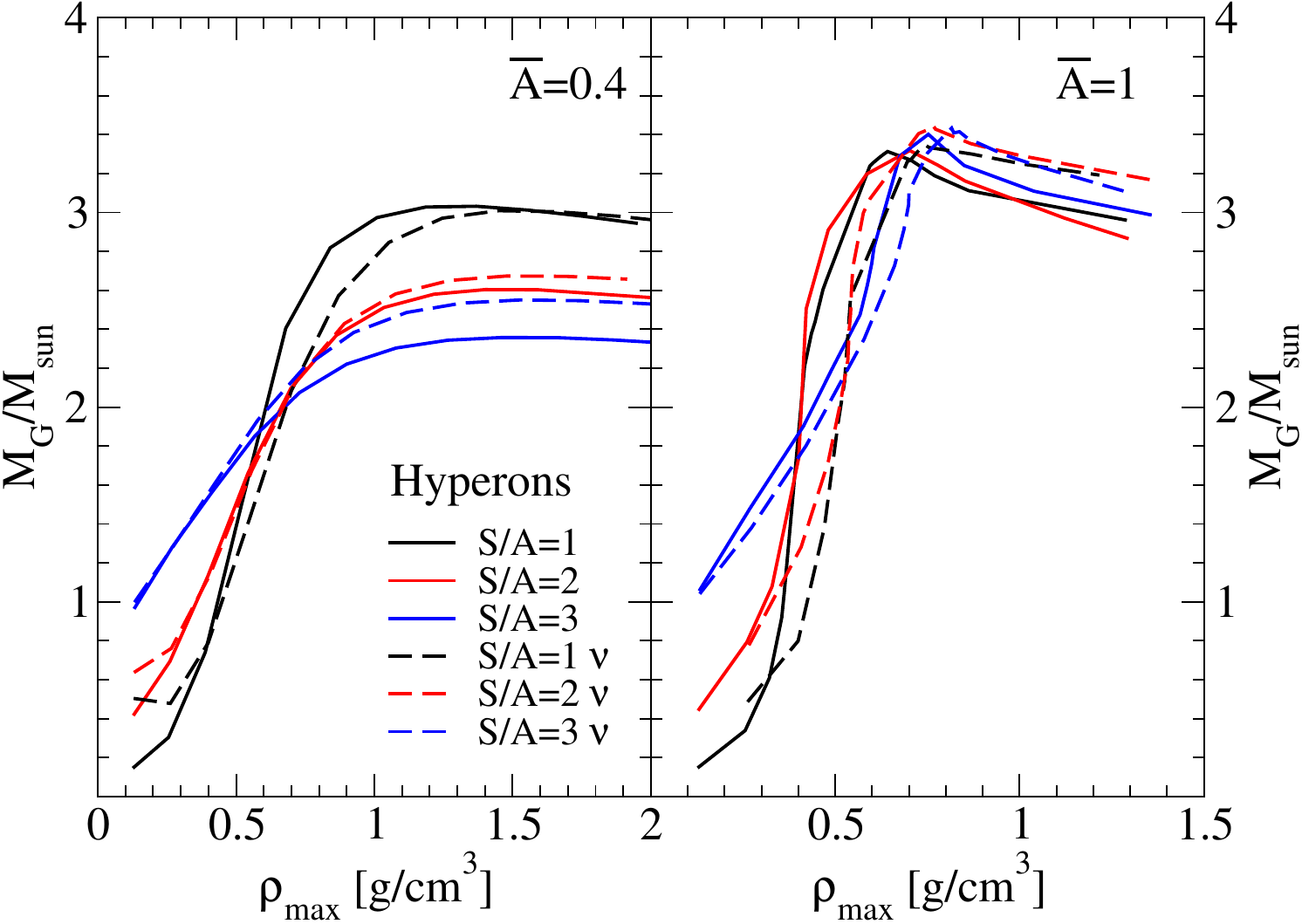}
\caption{The same of Fig.\ \ref{fig7} but for hyperonic matter. }
\label{fig8}
\end{figure}
In Fig.\ \ref{fig8} we present the same calculations as in Fig.\ \ref{fig7}, but for hyperonic matter. As in the nucleonic case, the left panel, corresponding to $\overline{A}=0.4$, is qualitatively similar to the rigid rotation scenario (see right panel of Fig.\ \ref{fig6}). By contrast, the right panel exhibits a rather different behavior. Here, we again find that the maximum mass supported increases with $S/A$, both for neutrino-free and neutrino-trapped matter, although in the case of hyperonic matter these variations are relatively small.

An interesting aspect confirmed by these calculations is that, even when differential rotation is included, the presence of neutrinos in hyperonic matter allows the star to support a larger gravitational mass compared with the neutrino-free case.

We emphasise, however, that due to limitations of the RNS code, the results for $\overline{A}=1$ should be interpreted with caution; the true maximum mass for these sequences may be somewhat higher. Nonetheless, we consider the qualitative trends reported here to be reliable. 
\cite{kaplan14} reported similar calculations for the LS220 and HShen EOSs, considering both zero-temperature and finite-temperature isothermal configurations with either cold or hot envelopes. Our results are again consistent with those obtained for EOSs with cold envelopes for both $\overline{A}=0.4$ and $\overline{A}=1$. 
In the case of hyperonic matter the scenario for $\overline{A}=1$ does not show a monotonic behavior of the maximum masses across the various models although the differences are very small. In the hyperonic case, the overall behavior is determined by a complex interplay between the softening (stiffening) of the EOS induced by the onset of hyperons in neutrino free (trapped) matter and the extreme centrifugal support provided by differential rotation. 

\begin{table}
\centering
\begin{tabular}{ccccccc}
\hline 
 EOS & $M_G$ & $M_B$ & $\rho_{max} $  & $R_e$  & $\Omega_c$ \\ 
     & [$M_{\odot}$] & [$M_{\odot}$] & $10^{15}$ [g/cm$^3$] & [km] & $10^3 \ {\rm rad} \ {\rm s}^{-1}$ \\ 
\hline
N T=0   & 3.372 & 3.915 & 0.879 &  17.877 & 12.667 \\  
N S/A=1 & 3.273 & 3.822 & 1.123 &  15.620 & 14.749 \\
N S/A=2 & 3.248 & 3.705 & 1.328 &  18.113 & 13.541 \\
N S/A=3 & 3.064 & 3.374 & 1.589 &  19.071 & 10.833 \\
\hline
N$\nu$  S/A=1 & 3.185 & 3.625 &  1.486 &  16.033 & 14.247 \\
N$\nu$  S/A=2 & 3.080 & 3.443 &  1.533 &  17.065 & 12.417 \\
N$\nu$  S/A=3 & 2.934 & 3.191 &  1.610 &  20.089 & 9.817 \\
\hline
Y T=0   & 3.018 & 3.452 & 1.124 &  17.154 & 11.061 \\ 
Y S/A=1 & 3.032 & 3.464 & 1.368 &  18.832 & 12.013 \\
Y S/A=2 & 2.604 & 2.891 & 1.401 &  18.939 &  9.563 \\
Y S/A=3 & 2.358 & 2.532 & 1.459 &  21.004 &  7.348 \\
\hline
Y$\nu$  S/A=1 & 3.009 & 3.393 & 1.449 &  18.319 & 11.614 \\
Y$\nu$  S/A=2 & 2.674 & 2.951 & 1.478 &  17.331 &  9.578 \\
Y$\nu$  S/A=3 & 2.551 & 2.734 & 1.528 &  20.959 &  7.766 \\
\hline
\end{tabular}
\caption{Maximum gravitational mass (first colum), baryonic mass (second column) configurations for various values of the entropy per baryon and including or not neutrino trapping. In the third and fourth columns we report the corresponding values of $\rho_{max}$ and $R_e$ while in the last column we give the value of central angular velocity $\Omega_c$.  
Results are shown for $\overline{A}=0.4$.  }
\label{tab3}
\end{table}


\begin{table*}
\centering
\begin{tabular}{ccccccc}
\hline 
 EOS & $M_G$ & $M_B$ & $\rho_{max} $  & $R_e$ & $\Omega_c$ \\ 
     & [$M_{\odot}$] & [$M_{\odot}$] & $10^{15}$ [g/cm$^3$] & [km] & [$10^3 \ {\rm rad} \ {\rm s}^{-1}$] \\ 
\hline
N  T=0   & 3.340 & 3.843 & 0.728 &  17.169 & 26.623 \\
N  S/A=1 & 3.410 & 3.912 & 0.754 &  17.242 & 27.437 \\
N  S/A=2 & 3.605 & 4.048 & 0.691 &  19.130 & 24.073 \\
N  S/A=3 & 3.724 & 4.061 & 0.620 &  22.331 & 18.265 \\
\hline
N$\nu$   S/A=1 & 3.338 & 3.743 &  0.799 &  17.552 & 25.437 \\
N$\nu$   S/A=2 & 3.517 & 3.883 &  0.767 &  18.998 & 23.710 \\
N$\nu$   S/A=3 & 3.609 & 3.897 &  0.736 &  21.431 & 19.156\\
\hline
Y T=0   & 3.274 & 3.737 & 0.829 &  17.744 & 26.071\\
Y S/A=1 & 3.313 & 3.746 & 0.642 &  18.599 & 22.046\\
Y S/A=2 & 3.318 & 3.686 & 0.702 &  19.278 & 20.736\\
Y S/A=3 & 3.402 & 3.679 & 0.752 &  21.697 & 17.900\\
\hline
Y$\nu$  S/A=1 & 3.345 & 3.733 & 0.745 &  18.260 & 23.620 \\
Y$\nu$  S/A=2 & 3.446 & 3.788 & 0.765 &  19.352 & 22.280 \\
Y$\nu$  S/A=3 & 3.432 & 3.685 & 0.815 &  21.402 & 18.644 \\
\hline
\end{tabular}
\caption{The same of Tab. \ref{tab3} but for $\overline{A}=1$.  }
\label{tab4}
\end{table*}

A summary of the maximum gravitational and baryonic mass configurations, corresponding central maximum energy density, equatorial radius and central angular velocity is reported in Tabs. \ref{tab3} and \ref{tab4} for $\bar{A}=0.4$ and for $\bar{A}=1$, respectively. 

\subsection{Remnant evolution}

Finally, we aim to explore how one can gain insights into the evolution of a remnant formed after a BNSM or a CCSNe. This evolution can be conveniently represented in a plane in which the gravitational mass, $M_G$, is plotted as a function of the baryonic mass, $M_B$. In a first approximation, the evolution of a remnant proceeds along a straight line at constant $M_B$. This is, however, not entirely accurate, since some mass is ejected during and after the formation of the remnant, which can be of the order of $\sim 0.05$–$0.1 \ M_\odot$ \citep{Bernuzzi20,endrizzi18,Radice2018} 
Accordingly, the trajectory followed by the initial remnant in the $(M_B, M_G)$ plane would be more complex. For the purposes of this discussion, we initially assume that the ejected mass is negligible, and subsequently comment on the more general and realistic scenario.

The idea is to illustrate a series of sequences representing a quasi-adiabatic evolution of the remnant at specific times, as divided into the following stages: \\

1) Immediately after the remnant’s formation, when neutrinos are trapped in the hot matter and the remnant rotates differentially; during this stage we fix $S/A = 2$. We refer to this kind of sequence as $\mathcal{A}$-type. \\
2) After a time interval of $\sim 0.1$–$0.5 \ \rm{s}$, differential rotation is lost and the star approaches a rigid rotation profile. 
During this stage, we consider two possible scenarios: in the first one neutrinos are still trapped and the matter remains hot (hereafter sequences $\mathcal{B}$); in the second scenario neutrinos escape from the star, producing a decrease of entropy and the  cooling of the star (hereafter sequences $\mathcal{C}$). We note that the two scenarios are not necessarily mutually exclusive. 
A transition $\mathcal{A}\rightarrow\mathcal{B}\rightarrow\mathcal{C}$ may indeed take splace.  \\ 
3) The final stage corresponds to the loss of rigid rotation and the cooling of the core; this physical situation is described by cold, deleptonised matter, i.e., a solution of the TOV equations (hereafter sequences $\mathcal{D}$).

It is implicit that each stage of the star’s evolution is governed by a specific EOS and its associated composition. Therefore, it is of interest to study how the remnant evolution may change according to different assumptions regarding matter composition, namely the presence of trapped neutrinos and/or the possible formation of hyperons.

We emphasise that, in our analysis, the timescales associated with the loss of differential rotation and with neutrino escape are crucial for describing the full evolution of a remnant. Unfortunately, these timescales, particularly that governing the loss of differential rotation, remain poorly known due to the complexity of the simultaneous processes involved  \citep{kiuchi18,Fuijib20,marg22,Radice23}.
Summarizing, according to the present uncertainties, in our scheme we discuss two possibile evolution paths: $\mathcal{A}\rightarrow\mathcal{B}\rightarrow\mathcal{C}\rightarrow\mathcal{D}$ and $\mathcal{A}\rightarrow\mathcal{C}\rightarrow\mathcal{D}$.   

In the following we focus on the case of a remnant formed after a BNSM. The case of a PNS can be clearly described using the very same methodology. The main differences between the two cases are that first, the remnant of CCSN is in general less massive than the corresponding one for a BNSM and, second, the rotational frequency for a PNS is much reduced comparing to the BNSM's case. This last aspect facilitates the dissipation of angular momentum as differential rotation is lost. (see discussion below).

\begin{figure*}[!ht]
\begin{centering}
\includegraphics[scale=0.3]{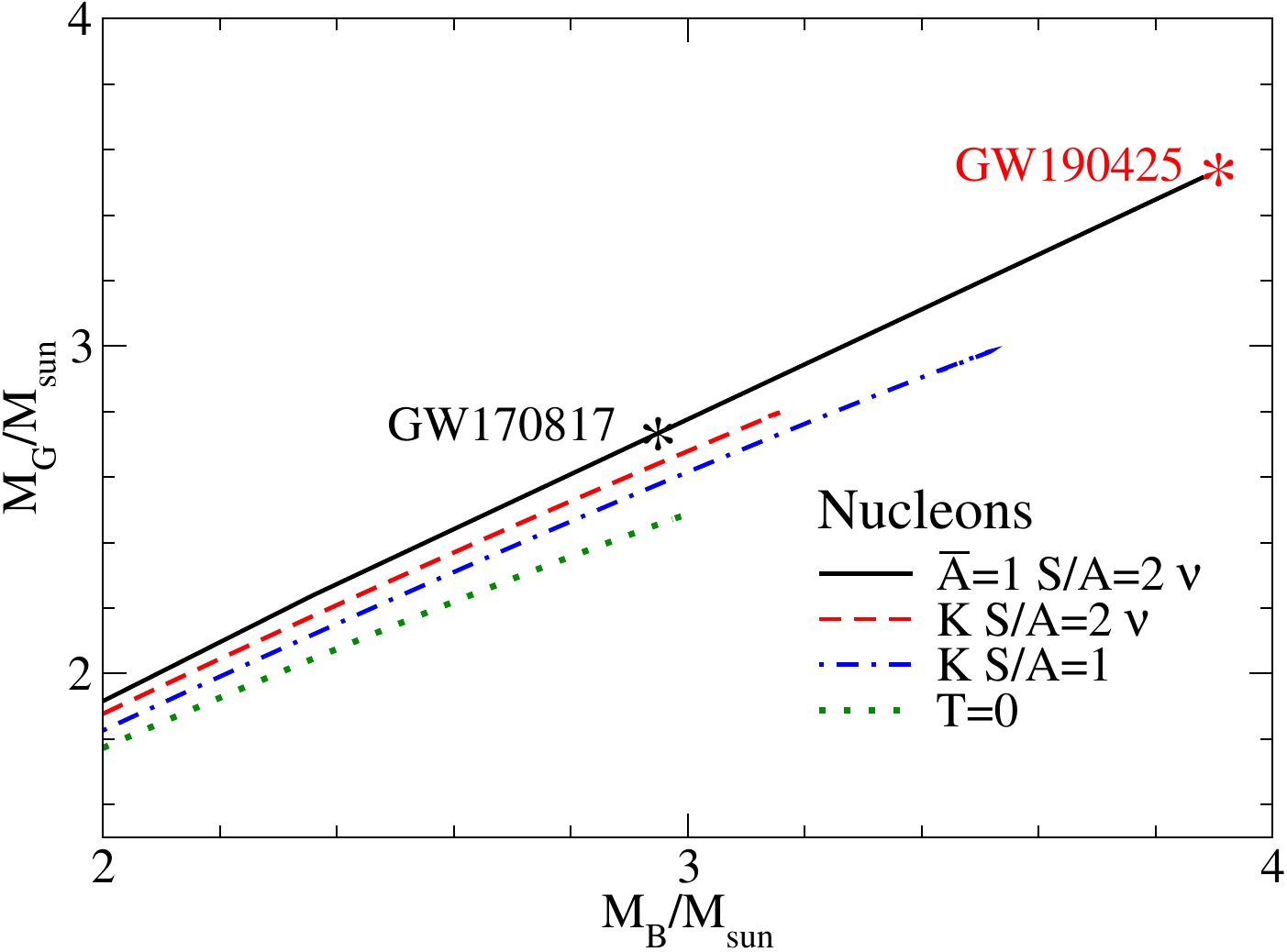}
\hspace{1cm}
\includegraphics[scale=0.3]{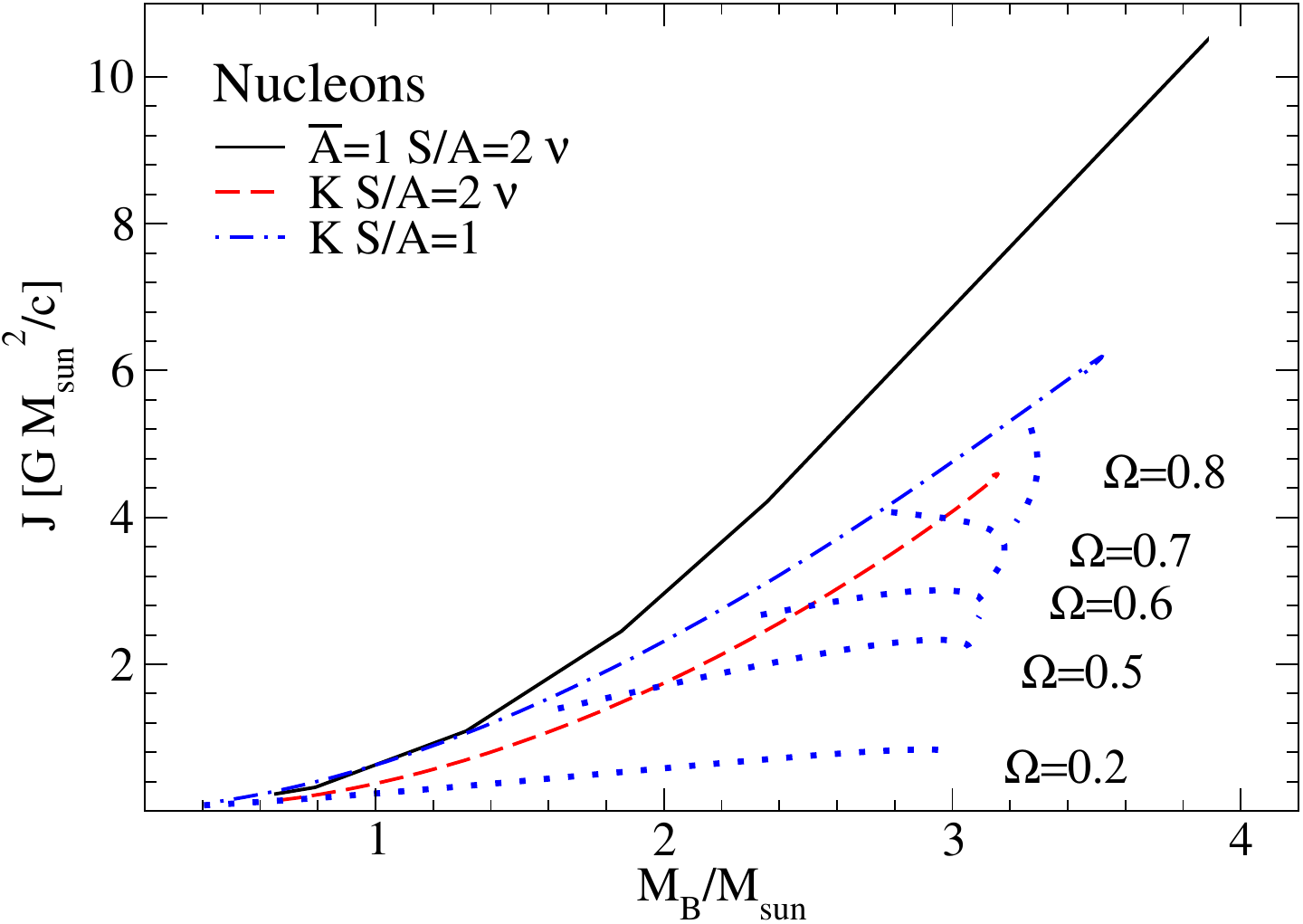}
\caption{In the left panel we report the remnants evolution in the plane ($M_B-M_G$) in the case of pure nucleonic matter. The black solid line show the sequence of differentially rotating stars considering a rotational profile with $\overline{A}=1$, $S/A=2$ and $\nu$-trapped. The red dashed line is the sequence of rigid rotating stars at mass-shed limit with $S/A=2$ and $\nu$-trapped. The blue dot-dashed line is the sequence of rigid rotating stars at mass shed limit with $S/A=1$ and $\nu$-free matter. The green dotted line stands for the sequence of cold nonrotating neutron stars.  
In the right panel we show the angular momentum $J$ in units of $G M_\odot^2/c$ corresponding to the same equilibrium sequences reported in the left panel. The blue dotted lines are sequences of rigidly rotating stars with angular velocity $\Omega<\Omega_K$. 
Note $\Omega$ is reported in units of $10^{4} \frac{\rm rad}{\rm s}$. The black and red asterisks mark the gravitational mass of the two BNSM events GW170817 and GW190425 respectively. Note that the resulting baryonic mass is a prediction of our model. }
\label{fig9}
\end{centering}
\end{figure*}

In the left panel of Fig. \ref{fig9}, we show the gravitational mass as a function of the baryonic mass for the various evolutionary sequences described above, in the case of pure nucleonic matter. The black line corresponds to the type-$\mathcal{A}$ sequence, with $\overline{A}=1$ and $S/A=2$. The red dashed and blue dash-dotted lines represent the type-$\mathcal{B}$ and type-$\mathcal{C}$ sequences, corresponding to $S/A=2$ with neutrino trapping and $S/A=1$ neutrino-free matter, respectively. Finally, the green dotted line indicates the type-$\mathcal{D}$ sequence, representing the final stage of the evolution.
As discussed above, the neutron stars evolution in this plane follows approximately straight lines at constant $M_B$. From the previous considerations, it is evident that if $M_B < M_B^{\rm max}(T=0)$, the model can in principle describes the formation of a neutron star with gravitational mass $M_G \le 2.484 \ M_\odot$, where $2.484 \ M_\odot$ corresponds to the maximum mass of the cold neutron star sequence. In this scenario (and for the specific values of parameters and EOS adopted), one may also conclude that the precise timescales for the loss of differential rotation and neutrino escape are not critical: starting from the upper sequence, there is always a stable configuration along the sequences below that allows the remnant to reach equilibrium. However, for this picture to hold, it is necessary that the transition from a differentially rotating configuration to a rigidly rotating one at the same $M_B$ dissipates the required amount of angular momentum.

In the right panel of Fig. \ref{fig9}, we show the angular momentum $J$ as a function of the baryonic mass for the same sequences displayed in the left panel. Inspection of these curves indicates that, in order to evolve from type-$\mathcal{A}$ to type-$\mathcal{B}$ configurations, some angular momentum $\Delta J$, given by the difference between the corresponding values at the same $M_B$, must be dissipated.

\begin{figure*}[!ht]
\begin{centering}
\includegraphics[scale=0.3]{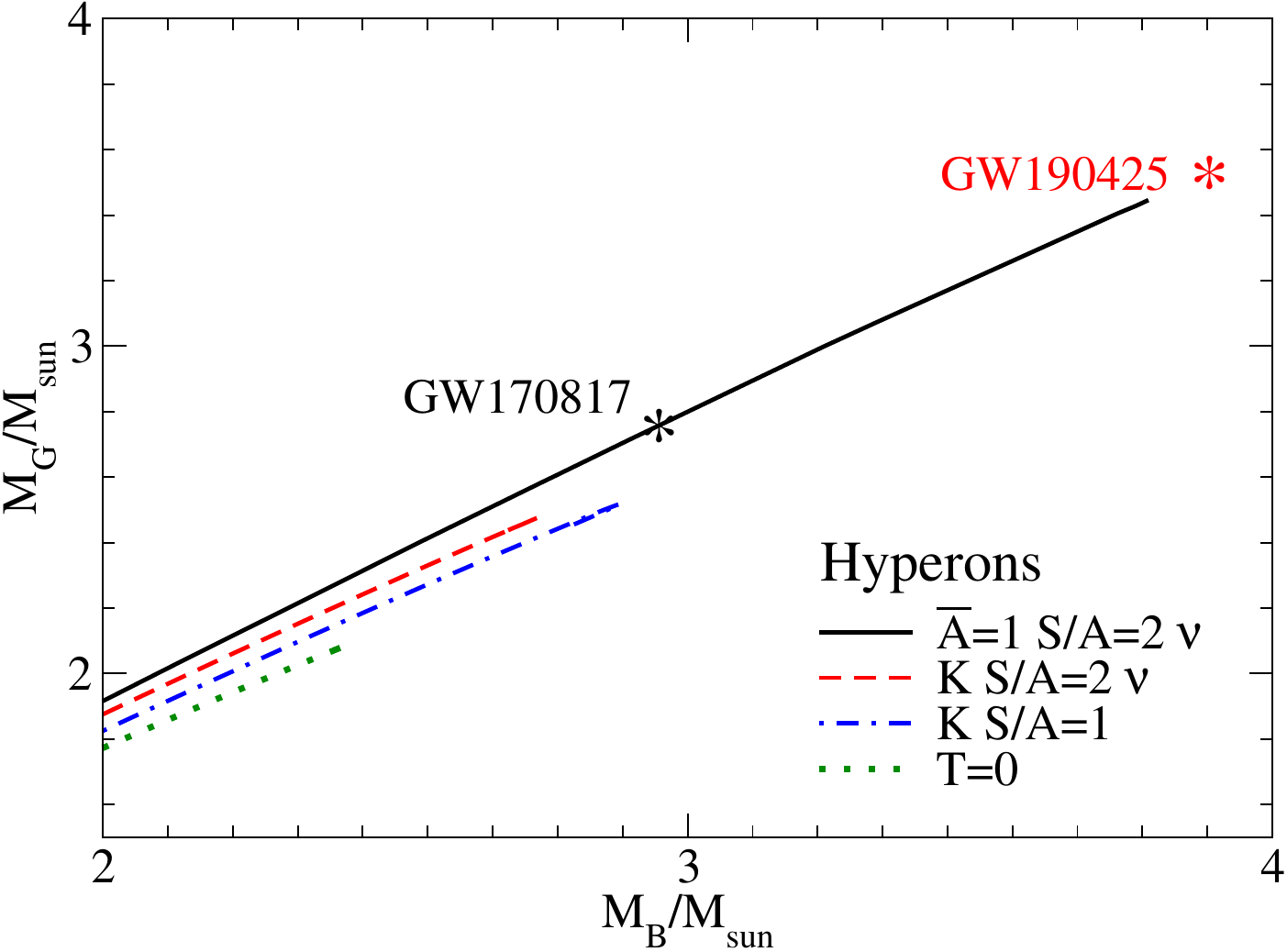}
\hspace{1cm}
\includegraphics[scale=0.3]{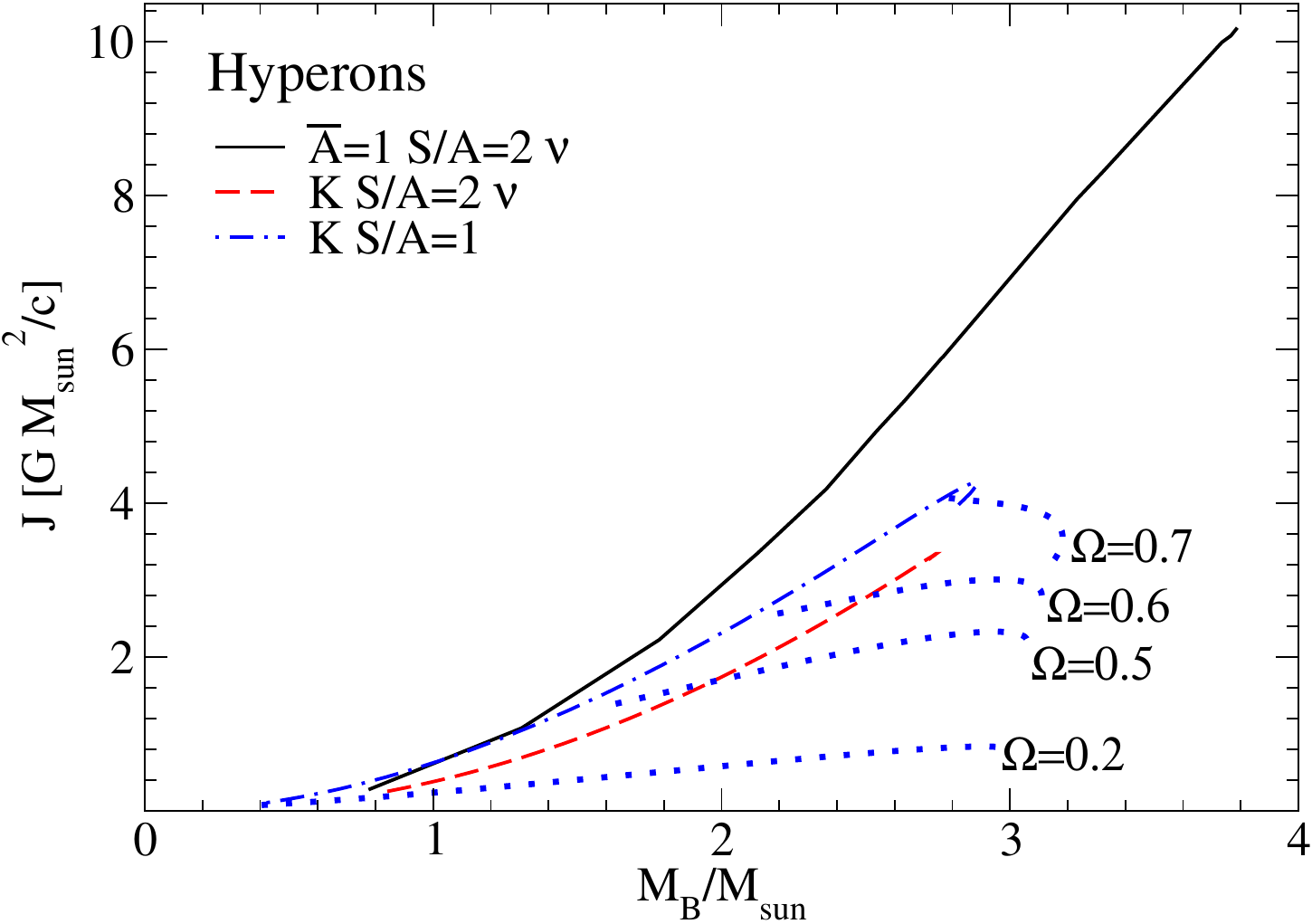}
\caption{The same of Fig.\ \ref{fig9} but for hyperonic matter. }
\label{fig10}
\end{centering}
\end{figure*}

Moreover, the transition $\mathcal{B} \rightarrow \mathcal{C}$ cannot occur, as it would require an unphysical spin-up of the system. However, slower rigidly rotating stars with $\Omega < \Omega_K$ (blue dotted sequences) that have smaller angular momenta than those of the $\mathcal{C}$ sequence can again be connected with the $\mathcal{B}$ configurations. In Fig. \ref{fig9}, the angular velocity $\Omega$ is reported in units of $10^4 , {\rm rad/s}$. Interestingly, for configurations at the crossing points between the blue dotted and red dashed sequences, angular momentum is effectively conserved during the transition.

Furthermore, if one considers a direct transition between the $\mathcal{A}$ and $\mathcal{C}$ sequences-corresponding to the case where differential rotation and neutrino escape occur on the same timescale-no such restriction exists, and a direct connection is always possible for all configurations. Note that, for clarity, the blue dotted sequences are not shown in the left panel of Fig. \ref{fig9}.

The black and red asterisks in Fig. \ref{fig9} indicate the gravitational masses of the two BNSM events GW170817 ($M_G \sim 2.72-2.76 \ M_\odot$) and GW190425 ($M_G \sim 3.3-3.7 \ M_\odot$), respectively, corresponding to the midpoint of the measured uncertainty intervals: $2.74 \ M_\odot$ for GW170817 and $3.50 \ M_\odot$ for GW190425. The corresponding baryonic masses are predicted by our model: for GW170817, using both nucleonic and hyperonic EOSs, and for GW190425, using the nucleonic EOS only.

According to our scheme, GW190425 likely underwent a prompt collapse to black-hole, whereas GW170817 formed a long-lived remnant. For GW170817, the angular momentum that must be dissipated for the transition $\mathcal{A} \rightarrow \mathcal{C}$ is $\Delta J \sim 2.14 \ G M_\odot^2 / c$, while for $\mathcal{A} \rightarrow \mathcal{B}$ it is $\Delta J \sim 2.88 \ G M_\odot^2 / c$. These values slightly  increase  if mass ejection is taken into account.

In our simplified model, the fate of GW170817 depends sensitively on the total mass of the system. If the system has the upper measured mass and mass ejection is neglected, a supramassive neutron star is formed, which would eventually collapse to a black hole. However, if at least $\sim 0.006 \ M_\odot$ of matter is ejected, a stable neutron star with $M_G \sim 2.484 \ M_\odot$ (the maximum mass supported by the TOV solution) could form after the loss of rigid rotation. The likelihood of a stable remnant increases if the total mass of the system is at the lower end of the measured interval. 

We have also tried to estimate the amount of mass ejecta required to support the angular momentum loss by using the quasiuniversal relations reported in \cite{Camilletti24}. The $\Delta J$'s would correspond to secular mass ejecta of $0.23\ M_\odot$ and of $0.32\ M_\odot$ respectively for the transitions $\mathcal{A} \rightarrow \mathcal{C}$ and $\mathcal{A} \rightarrow \mathcal{B}$. Being the latters way too high, this scenario assumes that GW170817 collapsed before reaching solid body rotation.

We additionally caution that this analysis does not account for possible secular instabilities that could affect the rigidly rotating neutron stars formed after differential rotation is lost. Similarly, the stability of our differentially rotating configurations is only assessed qualitatively; such instabilities could prevent the formation of a stable neutron star.

Nevertheless, our results, particularly regarding the stage of differential rotation, are broadly consistent with several full general relativistic simulations of GW170817 and GW190425 \citep{shib17,Bernuzzi20,Nedora21,Camilletti22,dudi22,raa21,figura21}. These studies indicate that, for various EOSs and both equal-mass and slightly asymmetric mergers, GW170817 did not undergo prompt collapse, forming instead a massive remnant neutron star. We emphasize that numerical simulations cannot cover evolution on timescales of several seconds, so our simplified framework provides a complementary tool for understanding the long-term evolution of merger remnants, albeit approximately.  

Let us now consider the analogous evolution scheme for hyperonic matter. In Fig. \ref{fig10}, we show the same sequences as in Fig. \ref{fig9}, but using our hyperonic EOSs. Some notable differences arise due to the softening of the EOS induced by the presence of hyperons. Our hyperonic model agrees with the nucleonic case in predicting that GW190425 likely underwent a prompt collapse, and that GW170817 initially formed a remnant. However, in the hyperonic scenario, the remnant of GW170817 collapses to a black hole as soon as differential rotation is lost; no supramassive neutron star is supported. Consequently, the lifetime of the GW170817 remnant is expected to be significantly shorter under this EOS. Based on this consideration, our analysis favors the hyperonic model over the nucleonic one.

We also note that using smaller values of the electron lepton fraction, $Y_{L_e}$, in neutrino-trapped matter the maximum baryonic and gravitational masses of the sequences shown in Figs. \ref{fig9} and \ref{fig10} slightly increases; however the overall qualitative scenario remains unchanged. Recent numerical simulations \citep{Radice23} indicate $Y_{L_e} \sim 0.05-0.1$ as realistic values, based on improved neutrino transport treatments. 

Overall, the simple scheme presented here provides a computationally inexpensive and intuitive way to gain insights into the long-term evolution of a remnant formed in a BNSM event or a CCSNe. Nevertheless, to obtain more definitive and quantitative results, fully dynamical numerical simulations remain indispensable in this field.

\section{Conclusions}
\label{conclusions}

In this work, we have employed cold and finite-temperature EOSs based on relativistic mean-field theory to analyze the structure of static as well as rotating neutron stars. Our EOSs predict static neutron star properties consistent with current observational constraints. Specifically, the maximum mass predicted by our models exceeds current estimates, while the tidal deformability for a $1.4 \ M_\odot$ neutron star lies within the limits inferred from the GW170817 event.

We have studied thermal effects in both static and rigidly rotating neutron stars, considering isoentropic EOSs and accounting for the possible trapping of neutrinos. We found that thermal effects in neutrino-free matter depend strongly on the presence of hyperons. In purely nucleonic matter, thermal effects enhance the maximum gravitational mass, whereas in hyperonic matter, they reduce it. This behavior arises because increasing $S/A$ promotes the thermal production of hyperons, which softens the EOS and lowers the maximum mass. In neutrino-trapped matter, the maximum mass is smaller than in neutrino-free matter for nucleonic matter, but larger for hyperonic matter. This trend results from the chemical equilibrium conditions among hadrons, as detailed in Section 2. Similar behaviors are observed for rigidly rotating stars; here the angular velocity corresponding to the Keplerian limit decreases with increasing $S/A$. Our results are in agreement with findings reported by other groups using different EOS models.

We then focused on differentially rotating neutron stars, employing the so-called $J$-constant law to describe the rotation profile. Two degrees of differential rotation were considered, parametrized by $\bar{A}=0.4$ (low degree) and $\bar{A}=1$ (high degree). For $\bar{A}=0.4$, the behavior is qualitatively similar to the rigid rotation case, with equilibrium solutions maintaining a spheroidal shape. In contrast, for $\bar{A}=1$, thermal effects support larger neutron star masses as $S/A$ increases, independently of hyperon content, and neutrino trapping. For these highly differentially rotating models, equilibrium structures become toroidal at large densities near the maximum supported mass. This leads to two important consequences: (1) the maximum energy density is shifted away from the stellar center, and (2) the central angular velocity becomes very high, providing strong centrifugal support.

Finally, we used these equilibrium sequences to explore the adiabatic evolution of remnants formed in BNSM or CCSN events. We analyzed this evolution in the $(M_B-M_G)$ plane, assuming that the baryonic mass ejected during formation is negligible. Applying the model to GW170817 and GW190425, we found that the dissipation of angular momentum plays a crucial role in determining remnant evolution. In addition, the presence of hyperons would lead to an earlier black hole collapse of the GW170817 remnant compared to the purely nucleonic case, where a longer-lived evolution is possible and collapse may occur on longer timescales. 
Our model indicates that a stable neutron star could potentially form in GW170817, although this scenario seems to be  disfavored by our estimate of the mass ejection required to dissipate the necessary angular momentum.
For GW190425, all considered models predict a prompt collapse. 
We finally note that in the present work we have neglected the effects of magnetic fields. The latter are known to act as efficient torque, promoting the transition to rigid rotation via magnetic braking and the transport of angular momentum within the star \citep{duez06,Bocquet95}. The inclusion and detailed study of these effects are left for future work.



\section*{Acknowledgement}  


SB acknowledges funding from the EU Horizon under ERC Consolidator Grant,
no. InspiReM-101043372 and from the DFG-NSF Physics: Multimessenger Astronomy of Neutron Star 
Mergers with Numerical Relativity (DFG BE 6301/8-1 no.~563106943).
DR acknowledges support from the Sloan Foundation, from the Department of
Energy, Office of Science, Division of Nuclear Physics under Award Number
DE-SC0024388, from the National Science Foundation under Grants No.
PHY-2020275, PHY-2116686, PHY-2407681, and PHY-2512802.



\end{document}